\documentclass[
journal=jpccck, 
manuscript=article,
]{achemso}

\usepackage{amsmath,amssymb,bm}
\usepackage{graphicx}
\usepackage{color} 
\usepackage{hyperref}
\usepackage{mathrsfs} 
\usepackage{relsize} 

\newcommand*\diff{\mathrm{d}^3}

\SectionNumbersOn

\author{A. H. Davoody}
\email{davoody@wisc.edu}
\affiliation{Department of Electrical and Computer Engineering, University of Wisconsin--Madison, Madison, WI 53706-1691, USA}
\author{F. Karimi}
\affiliation{Department of Electrical and Computer Engineering, University of Wisconsin--Madison, Madison, WI 53706-1691, USA}
\author{M. S. Arnold}
\affiliation{Department of Materials Science and Engineering, University of Wisconsin--Madison, Madison, WI 53706-1691, USA}
\author{I. Knezevic}
\email{iknezevic@wisc.edu}
\phone{(608) 890-3383}
\affiliation{Department of Electrical and Computer Engineering, University of Wisconsin--Madison, Madison, WI 53706-1691, USA}

\title{Theory of Exciton Energy Transfer in Carbon Nanotube Composites}

\begin{document}

\setlength{\fboxrule}{0 pt}



\begin{abstract}
We compute the exciton transfer (ET) rate between semiconducting single-wall carbon nanotubes (SWNTs). We show that the main reasons for the wide range of measured ET rates reported in the literature are 1) exciton confinement in local quantum wells stemming from disorder in the environment and 2) exciton thermalization between dark and bright states due to intratube scattering. The SWNT excitonic states are calculated by solving the Bethe-Salpeter equation using tight-binding basis functions. The ET rates due to intertube Coulomb interaction are computed via Fermi's golden rule. In pristine samples, the ET rate between parallel (bundled) SWNTs of similar chirality is very high ($\sim 10^{14}\;\text{s}^{-1}$) while the ET rate for dissimilar or nonparallel tubes is considerably lower ($\sim 10^{12}\;\text{s}^{-1}$). Exciton confinement reduces the ET rate between same-chirality parallel SWNTs by two orders of magnitude, but has little effect otherwise. Consequently, the ET rate in most measurements will  be on the order of $ 10^{12}\;\text{s}^{-1}$, regardless of the tube relative orientation or chirality.  Exciton thermalization between bright and dark states further reduces the ET rate to about $10^{11}\;\text{s}^{-1}$. The ET rate also increases with increasing temperature and decreases with increasing dielectric constant of the surrounding medium. \end{abstract}

\maketitle


\section{Introduction}
Carbon nanotubes (CNTs) are quasi-one-dimensional materials with a unique set of optical and electronic properties \cite{Jariwala2013carbon,Arnold2013recent}. Today, there is considerable interest in semiconducting CNTs as the light-absorbing material in organic solar cells, owing to their tunable band gap, excellent carrier mobility, and chemical stability\cite{Jain2012polymer,Bindl2013enhansing,Ye2014semiconducting,Bindl2011semiconducting,Bindl2010dissociating,Ham2010evidence,Bindl2011efficiently}. Improving the efficiency of CNT-based photovoltaic devices is possible through understanding the dynamics of excitons in CNT composites.

While the intratube dynamics of excitons in CNTs have been studied extensively over the past decade \cite{Luer2009size, Perebeinos2005radiative, Spataru2005theory, Miyauchi2009radiative, Ostojic2004interband, Huang2004ultrafast, Graham2010exciton, Manzoni2005intersubband, Hagen2005exponential}, the intertube exciton dynamics remain  relatively unexplored, owing to the difficulties in sample preparation and measurements. The fluorescence from bundled single-wall carbon nanotube (SWNT) samples is quenched and the absorption spectra are broadened as a result of Coulomb interaction between SWNTs with various chiralities\cite{OConnell2001band}. The exciton lifetimes in isolated and bundled SWNTs were measured to be on the order of \(100 \mbox{ ps}\) and \(1-2 \mbox{ ps}\), respectively \cite{Huang2004ultrafast}, which underscores the importance of intertube Coulomb interactions in the dynamics of excitons in SWNT aggregates.

There have been a number of measurements of the exciton transfer (ET) rates in CNTs, but the reported rates differ widely, within two orders of magnitude \cite{Koyama2012ultrafast}. Pump-probe (PP) spectroscopy measurements have shown a time constant of about \( 0.37 \mbox{ ps}\) for the ET process from semiconducting SWNTs to metalic SWNTs \cite{Maeda2006gigantic}. Time-resolved photoluminescence (PL) spectroscopy found the time constants of \(0.9 \mbox{ ps}\) and \(0.5 \mbox{ ps}\) for ET from semiconducting SWNTs to metallic and semiconducting SWNTs, respectively \cite{Koyama2010ultrafast}. In another study, time-resolved PL spectroscopy has been used to measure ET between semiconducting SWNTs, with a time constant \(\tau \approx 70 \mbox{ ps}\) \cite{Berger2007temperature}. Qian \textit{et al.} used spatial high-resolution optical spectroscopy to estimate the time constant of ET between two nonparallel semiconducting SWNTs  as \(\tau \approx 0.5 \mbox{ ps}\) \cite{Qian2008exciton}. Using PP spectroscopy, the ET time constant between bundled SWNTs was measured to be \(\tau \approx 10 \mbox{ fs}\) for \(S_{11}\) excitons.
\bibnote{\(S_{ij}\) exciton corresponds to the electron transition from \(i\)-th highest valence to the \(j\)-th lowest conduction bands.}
\cite{Luer2010ultrafast} The same study estimated the \(S_{22}\) exciton transfer to be very slow due to momentum mismatch. Another study showed a long-range fast component (\(\tau \approx 0.3 \mbox{ ps}\)),  followed by a short-range slow component (\(\tau \approx 10 \mbox{ ps}\)) for the ET process in SWNT films \cite{Mehlenbacher2013photoexcitaion}.  Grechko \textit{et al.} used a diffusion-based model to explain their measurement of ET in bundled semiconducting SWNT samples \cite{Grechko2014diffusion}. They found the time constants of \(\tau \sim 0.2-0.4 \mbox{ ps}\) for ET between bundles of SWNTs and \(\tau \approx 7 \mbox{ ps}\) for ET within SWNT bundles. A more recent study by Mehlenbacher \textit{et al.}  revealed ultrafast \(S_{22}\) exciton transfer \cite{Mehlenbacher2015energy}.

Many difficulties inherent in experiments can be avoided in a theoretical study of the transfer process. However, to date there have been only two theoretical studies of exciton transfer between semiconducting SWNTs \cite{Wong2009ideal, Postupna2014photoinduced}. Wong \textit{et al.} showed that the ideal dipole approximation (known as the F\"orster theory) overestimates the exciton transfer rate by three orders of magnitude \cite{Wong2009ideal}. Postupna \textit{et al.} showed that exciton--phonon coupling could have a prominent effect on the exciton transfer process between (6,4) and (8,4) SWNTs \cite{Postupna2014photoinduced}. However, these studies did not account for some important parameters, such as the existence of low-lying optically dark excitonic states, chirality and diameter of donor and acceptor CNTs, temperature, confinement of excitons, screening due to the surrounding medium, and the interaction between various exciton subbands, all of which have been shown, experimentally and theoretically, to play an important role in exciton dynamics in CNTs \cite{Perebeinos2004scaling, Nugraha2010dielectric, Perebeinos2005radiative, Spataru2005theory, Mehlenbacher2013photoexcitaion}.

In this paper, we present a comprehensive theoretical analysis of Coulomb-mediated intertube exciton dynamics in SWNT composites, in which we pay attention to the complex structure of excitonic dispersions, exciton confinement, screening due to surrounding media, and temperature dependence of the ET rate. We solve the Bethe-Salpeter equation in the GW approximation in the basis of single-particle states obtained from nearest-neighbor tight binding in order to calculate the exciton dispersions and wave functions. We calculate the intertube exciton transfer rate due to the Coulomb interaction between SWNTs of different chiralities and orientations. For the sake of brevity,  in the rest of this paper, we refer to single-wall carbon nanotubes simply as carbon nanotubes unless otherwise noted.

We show that momentum conservation plays an important role in determining the ET rate between parallel CNTs of different chirality. While the ET rate between similar-chirality bundled parallel tubes in pristine samples is $\sim 10^{14}\;\text{s}^{-1}$  \cite{Mehlenbacher2016ultrafast}, much higher than between misoriented or different-chirality CNTs ($\sim 10^{12}\;\text{s}^{-1}$), exciton confinement due to disorder strongly reduces the ET rate between parallel tubes of similar chirality, but has little effect on the ET rate otherwise. Consequently, the ET rate dependence on the orientation of donor and acceptor CNT is not as prominent as predicted previously and the ET rate is instead expected to be isotropic and $~10^{12}\;\text{s}^{-1}$ in most experiments \cite{Mehlenbacher2016ultrafast}. Moreover, the exciton transfer rate drops by about one order of magnitude if intratube exciton scattering between bright and dark excitonic states is allowed. Our study shows that the transfer from \(S_{22}\) to \(S_{11}\) excitonic states happens with the same rate as the transfer process between same-subband states (\(S_{11} \rightarrow S_{11}\) and $S_{22} \rightarrow S_{22}$). The ET rate increases with increasing temperature. We also show that the screening of the Coulomb interaction by the surrounding medium reduces the transfer rate by changing the exciton wave function and energy dispersion, as well as by reducing the Coulomb coupling between the donor and acceptor CNTs.

The rest of this paper is organized as follows. In Sec. \ref{sec:excitons_in_CNTs}, we provide a summary of the physics of excitonic states in CNTs. We review the calculation of exciton wave functions by using the tight-binding states as the basis functions. In Sec. \ref{sec:resonance_energy_transfer_rate}, we introduce the formulation of excitonic energy transfer in CNTs. Section \ref{sec:results_and_discussion} shows the ET rate between various excitonic states of CNTs and discusses the effect of the above mentioned parameters. Section \ref{sec:summary} provides a summary of the results. Appendix \ref{sec:derivation_of_free_exciton_transfer_rate} shows a detailed derivation of the exciton transfer rates between very long CNTs.

\section{Excitons in carbon nanotubes} \label{sec:excitons_in_CNTs}
In this section, we provide an overview of excitonic states in CNTs. A detailed derivation of these formulas can be found in papers of Rohlfing \textit{et al.} \cite{Rohlfing2000electron} and Jiang \textit{et al.} \cite{Jiang2007chirality}

\subsection{Single-particle states}
\begin{figure}[!tbp]
  \centering
  \includegraphics[width=0.95\linewidth]{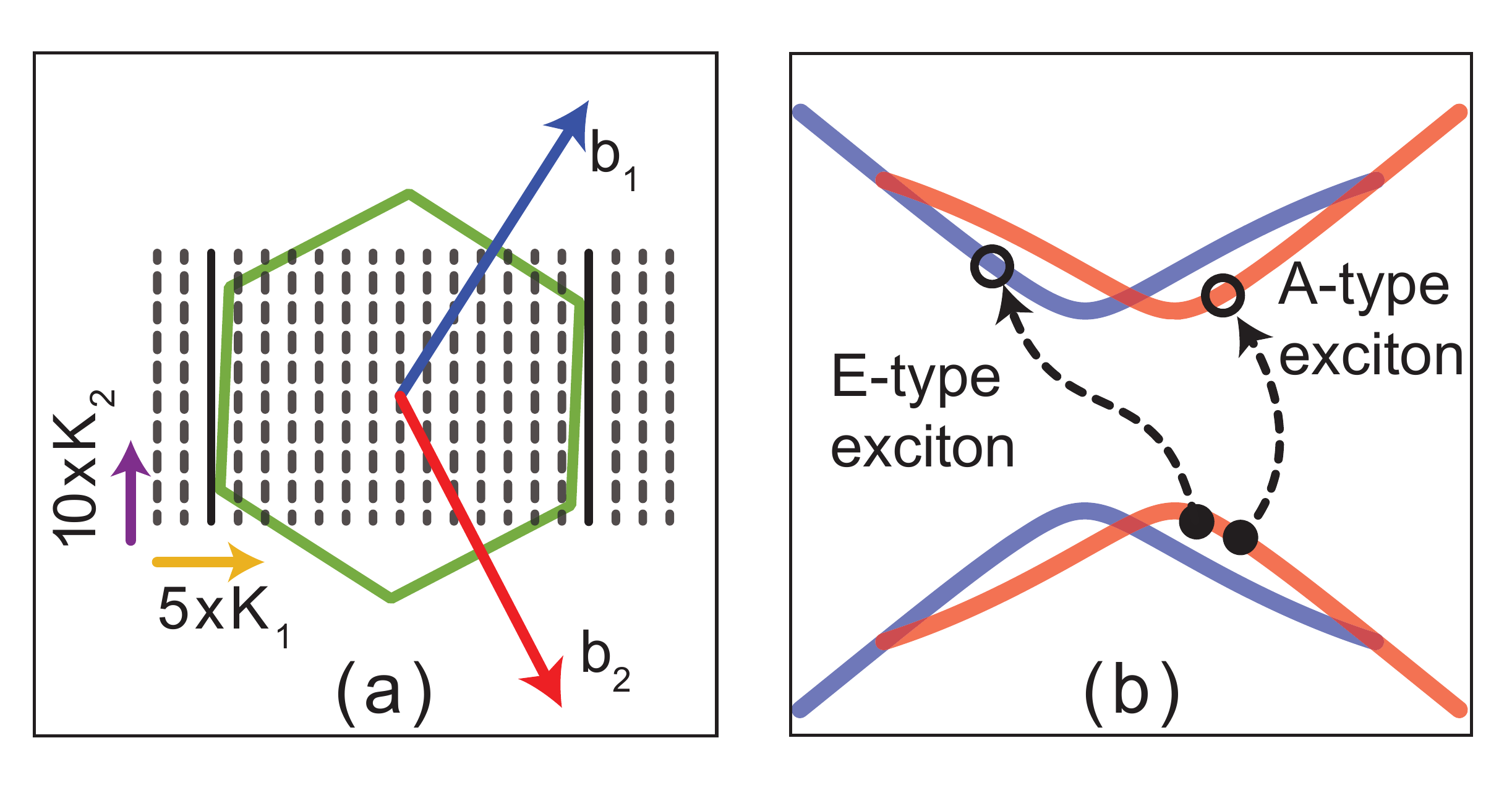}\\
  \caption{(a) Graphene Brillouin zone and the cutting lines of a CNT with (7,6) chirality. Bold solid lines represent the degenerate cutting lines that  pass by the \(K\) and \(K'\) points in the graphene Brillouin zone. (b) Schematic of excitation type based on the cutting line of electron and hole.}
  \label{fig:Brillouin_zone_and_exciton_type}
\end{figure}

For the purpose of calculating the CNT electronic structure, we can consider a CNT as a rolled graphene sheet. Within this picture, a CNT electronic structure is the same as for a graphene sheet. Using the tight binding (TB) method, the CNT wave function is \cite{Saito1998physical}
\begin{equation}
    \psi_{a\bm{k}}(\bm{r})=\frac{1}{\sqrt{N_u}}\sum_b\sum_u C_{ab}(\bm{k}) e^{i\bm{k}.\bm{R}_{ub}}\phi(\bm{r}-\bm{R}_{ub}),
\end{equation}
where \(u\) runs over all the \(N_u\) graphene unit cells, \(b=A,B\) runs over all the basis atoms in a graphene unit cell, and \(a\) is the band index. \(\phi(\bm{r})\) is the \(p_z\) orbital of carbon atom located at the origin. In order to satisfy the azimuthal symmetry of the wave function, the wave vector is limited to specific values known as cutting lines
\begin{equation}
    \bm{k} = \mu \bm{K}_1 + k \bm{K}_2/|\bm{K}_2|.
\end{equation}
Here, \(\mu\) is an integer, determining the cutting line. \(\bm{K}_1\) and \(\bm{K}_2\) are the reciprocal lattice vectors along the circumferential and axial directions, respectively (Figure \ref{fig:Brillouin_zone_and_exciton_type}a). It is noteworthy that there are always two degenerate cutting lines that pass by the two Dirac points (\(K\) and \(K'\)) in the graphene Brillouin zone.

\subsection{Excitonic states}
According to the Tamm-Dancoff approximation, the electron-hole excitation (exciton) wave function is a linear combination of free electron-free hole wave functions \cite{Fetter1971quantum}
\begin{equation} \label{eq:TammDancoffApproximation}
    |n\rangle=\sum_{\bm{k}_c,\bm{k}_v}A_n(\bm{k}_c,\bm{k}_v)\hat{u}^{\dagger}(\bm{k}_c)\hat{v}(\bm{k}_v)|\text{GS}\rangle.
\end{equation}
Here, \(A_n(\bm{k}_c,\bm{k}_v)\) is the expansion coefficient, \(\hat{u}^{\dagger}(\bm{k}_c)\) is the creation operator of an electron in the conduction band with wave vector \(\bm{k}_c\), and \(\hat{v}(\bm{k}_v)\) is the annihilation operator of an electron from the valence band with wave vector \(\bm{k}_v\). \(|\mbox{GS}\rangle\) is the system ground state, which corresponds to the valence band full and the conduction band empty of electrons. \(n\) represents all the quantum numbers.  We use the nearest-neighbor tight-binding single-particle wave functions as the basis functions. The expansion coefficients in the exciton wave function and the excitonic eigenenergies are calculated by solving the Bethe-Salpeter (BS) equation

\begin{equation}
    [E_c(\bm{k}_c)-E_v(\bm{k}_v)]A_n(\bm{k}_c,\bm{k}_v)+ \\
    \sum_{\bm{k}'_c,\bm{k}'_v}\mathcal{K}(\bm{k}_c,\bm{k}_v;\bm{k}'_c,\bm{k}'_v)A_n(\bm{k}'_c,\bm{k}'_v)= \Omega_n A_n(\bm{k}_c,\bm{k}_v),
\end{equation}

\noindent where \(E_c(\bm{k}_c)\) and \(E_v(\bm{k}_v)\) are the quasiparticle energies of electrons with wave vectors \(\bm{k}_c\) and \(\bm{k}_v\) in the conduction and valence bands, respectively. \(\Omega_n\) is the exciton energy. \(\mathcal{K}\) is the interaction kernel that describes the particle-particle interaction. Using the \emph{GW} approximation \cite{Strinati1988application}, we can divide the interaction kernel into the direct and exchange terms
\begin{subequations}
    \begin{equation}
        \mathcal{K}(\bm{k}_c,\bm{k}_v;\bm{k}'_c,\bm{k}'_v)= \\
        2\delta_\alpha \mathcal{K}^x(\bm{k}_c,\bm{k}_v;\bm{k}'_c,\bm{k}'_v)- \mathcal{K}^d(\bm{k}_c,\bm{k}_v;\bm{k}'_c,\bm{k}'_v),
    \end{equation}
    where the direct interaction depends on the screened Coulomb potential, \(w\),
    \begin{equation}
    \begin{split}
        \mathcal{K}^d(\bm{k}_c,\bm{k}_v;\bm{k}'_c,\bm{k}'_v)= & ~ W(\bm{k}_c,\bm{k}'_c;\bm{k}_v,\bm{k}'_v) \\
        = & \int \diff \bm{r} \, \diff \bm{r}' \psi^*_{\bm{k}_c}(\bm{r}')\psi_{\bm{k}'_c}(\bm{r}') w(\bm{r},\bm{r}';\omega=0) \psi_{\bm{k}_v}(\bm{r})\psi^*_{\bm{k}'_v}(\bm{r}),
    \end{split}
    \end{equation}
    and the exchange interaction is calculated using the bare Coulomb potential, \(v\),
    \begin{equation}
    \begin{split}
        \mathcal{K}^x(\bm{k}_c,\bm{k}_v;\bm{k}'_c,\bm{k}'_v)= & ~ V(\bm{k}_c,\bm{k}_v;\bm{k}'_c,\bm{k}'_v) \\
       = & \int \diff \bm{r} \, \diff \bm{r}' \psi^*_{\bm{k}_c}(\bm{r}')\psi_{\bm{k}_v}(\bm{r}') v(\bm{r},\bm{r}') \psi_{\bm{k}'_c}(\bm{r})\psi^*_{\bm{k}'_v}(\bm{r}).
    \end{split}
    \end{equation}
\end{subequations}
Here, \(\psi_{\bm{k}}\) is the quasiparticle (electron or hole) wave function. \(\alpha\) is the exciton spin, and \(\delta_{\alpha}=1\) for singlet excitons (\(\alpha=0\)) and \(\delta_{\alpha}=0\) for triplet excitons (\(\alpha=1\)).

The screened Coulomb interaction can be calculated using the random phase approximation \cite{Jiang2007chirality,Brener1975screened}

\begin{equation} \label{eq:screened_coulomb_interaction}
    W(a_1\bm{k}_1,a_2\bm{k}_2;a_3\bm{k}_3,a_4\bm{k}_4)=\frac{V(a_1\bm{k}_1,a_2\bm{k}_2;a_3\bm{k}_3,a_4\bm{k}_4)}{\kappa \epsilon^{\pi}_{r}(\bm{k}_1-\bm{k}_2,\omega=0)}.
\end{equation}
Here, \(a_i\) represents the conduction or valence band of graphene. $\kappa \epsilon^{\pi}_{r}$ is the relative dielectric permittivity of the tube: $\kappa$ is the static dielectric permittivity that accounts for the effect of core electrons and the surrounding environment, while the effect of \(\pi\)-bond electrons is considered in the dynamic relative dielectric function \(\epsilon^{\pi}_{r}(\bm{q},\omega)\), which is calculated through the Lindhard formula \cite{Brener1975screened}.

Using tight-binding single-particle wave functions, the interaction kernels are
\begin{equation} \label{eq:InteractionKernel}
    \begin{split}
        & \mathcal{K}^d(\bm{k}_c,\bm{k}_v;\bm{k}'_c,\bm{k}'_v)=\delta(\bm{k}'_c-\bm{k}_c,\bm{k}'_v-\bm{k}_v) \times \cdots \\
        & \hspace{1.5 cm} \cdots \sum_{b,b'}C_{cb}^*(\bm{k}_c)C_{cb}(\bm{k}'_c) C_{vb'}(\bm{k}_v)C_{vb'}^*(\bm{k}'_v) \frac{v_{b,b'}(\bm{k}_c-\bm{k}'_c)}{\kappa \epsilon^{\pi}_{r}(\bm{k}_c-\bm{k}'_c,0)}, \\
        &\mathcal{K}^x(\bm{k}_c,\bm{k}_v;\bm{k}'_c,\bm{k}'_v)=\delta(\bm{k}_v-\bm{k}_c,\bm{k}'_v-\bm{k}'_c) \times \cdots \\
        & \hspace{1.5 cm} \cdots \sum_{b,b'}C_{cb}^*(\bm{k}_c)C_{vb}(\bm{k}_v) C_{cb'}(\bm{k}'_c)C_{vb'}^*(\bm{k}'_v) v_{b,b'}(\bm{k}_c-\bm{k}_v),
    \end{split}
\end{equation}
where \(v_{b,b'}(\bm{q})\) is the Fourier transform of the overlap matrix between two \(p_z\) orbitals
\begin{equation}
    \begin{split}
        & v_{b,b'}(\bm{q})=\frac{1}{N_u}\sum_{u''}e^{i\bm{q}.(\bm{R}_{u''b'}-R_{0b})}I(\bm{R}_{u''b'}-\bm{R}_{0b}), \\
        & I(\bm{R}_{u'b'}-\bm{R}_{ub})= \int \diff \bm{r}\, \diff \bm{r}' |\phi(\bm{r}-\bm{R}_{ub})|^2 v(|\bm{r}-\bm{r}'|) |\phi(\bm{r}'-\bm{R}_{u'b'})|^2 \\
        & \hspace{2.6 cm} \approx \frac{U}{\sqrt{\left(\frac{4\pi\epsilon_0}{e^2}U |\bm{R}_{u''b'}-\bm{R}_{0b}|\right)^2+1}},
    \end{split}
\end{equation}

The last expression is the Ohno potential \cite{Ohno1964some}, where we take the parameter \(U = 11.3\) eV. Note that, due to the delta functions in Eq. (\ref{eq:InteractionKernel}), we can write the interaction kernels in terms of the center-of-mass and relative-motion wave vectors \bibnote{Note that in the electron-hole picture, the wave vector of excited electron in the valence band is negative of the wave vector of created hole: \(\bm{k}_h = -\bm{k}_v\).}
\begin{equation}
    \bm{K}=\frac{\bm{k}_c-\bm{k}_v}{2} \quad , \quad \bm{k}_r=\frac{\bm{k}_c+\bm{k}_v}{2}.
\end{equation}
Consequently, the center-of-mass wave vector is a good quantum number and the quantum number $n$ in the BS equation has two components: $n =  (s, \bm{K})$. $s$ is the quantum number analogous to the principal quantum number in a hydrogen atom. The excited state in Eq. (\ref{eq:TammDancoffApproximation}) is now be written as
\begin{equation} \label{eq:TammDancoffModified}
    |s,\bm{K}\rangle=\sum_{\bm{k}_r} A_s(\bm{K},\bm{k}_r)\hat{u}^{\dagger}(\bm{k}_r+\bm{K})\hat{v}(\bm{k}_r-\bm{K})|\mbox{GS}\rangle.
\end{equation}

\noindent Note that both the center-of-mass and relative-motion wave vectors are two-dimensional vectors consisting of two components. (i) A discrete circumferential part, which determines the electron and hole subband indices,
\begin{equation}
    \mathscr{M} = (\mu_c-\mu_v)/2 \quad , \quad \mu_r = (\mu_c+\mu_v)/2.
\end{equation}
\(\mathscr{M}\) determines the angular momentum of the exciton center of mass. (ii) A continuous part, parallel to the axis of the CNT, which determines how fast they are moving along the CNT,
\begin{equation}
    K = (k_c-k_v)/2 \quad , \quad k_r = (k_c+k_v)/2.
\end{equation}

\noindent If the exciton center-of-mass angular momentum is zero (\(\mathscr{M} = 0 \)) the exciton is called \(A\)-type. In an \(A\)-type exciton, the electron and the hole belong to the same cutting lines. On the other hand, if the electron and the hole belong the different cutting lines, the center-of-mass angular momentum is nonzero (\(\mathscr{M} \neq 0\)) and the exciton is called \(E\)-type (See Fig. \ref{fig:Brillouin_zone_and_exciton_type}b); \(E_{+}\) and \(E_{-}\) refer to excitons with positive and negative center-of-mass angular momenta, respectively.

For an \(A\)-type exciton (\(\mathscr{M}=0\)), it is easy to show that
\begin{subequations}
    \begin{equation}
        \mathcal{K}^d(\bm{k}_r,\bm{k}'_r;\bm{K})=\mathcal{K}^d(-\bm{k}_r,-\bm{k}'_r;\bm{K}),
    \end{equation}
    \begin{equation}
        \mathcal{K}^x(\bm{k}_r,\bm{k}'_r;\bm{K})=\mathcal{K}^x(-\bm{k}_r,-\bm{k}'_r;\bm{K}),
    \end{equation}
    \begin{equation}
        \mathcal{K}^x(\bm{k}_r,\bm{k}'_r;\bm{K})=\mathcal{K}^x(\bm{k}_r,-\bm{k}'_r;\bm{K}),
    \end{equation}
\end{subequations}

\noindent which results in symmetric (\(A_1\)) and antisymmetric (\(A_2\)) excitons \cite{Barros2006selection}

\begin{equation}
    A_1 \mbox{ exciton} \rightarrow A_s(\bm{K},\bm{k}_r) = - A_s(\bm{K},-\bm{k}_r),
\end{equation}
\begin{equation}
    A_2 \mbox{ exciton} \rightarrow A_s(\bm{K},\bm{k}_r) = + A_s(\bm{K},-\bm{k}_r).
\end{equation}

\noindent The wave function of an \(A_1\)(\(A_2\)) exciton are symmetric (antisymmetric) under \(C_2\) rotation around an axis perpendicular to the unwrapped graphene sheet. Also, the \(E\)-type excitons center-of-mass can rotate clockwise (\(E_+\)) or anticlockwise (\(E_-\) along the circumference of the CNT. Among the excitons discussed here, only the singlet \(A_2\) exciton is optically active (bright) and the rest are dark excitons.

\begin{figure}[!tbp]
  \centering
  \includegraphics[width=0.85\linewidth]{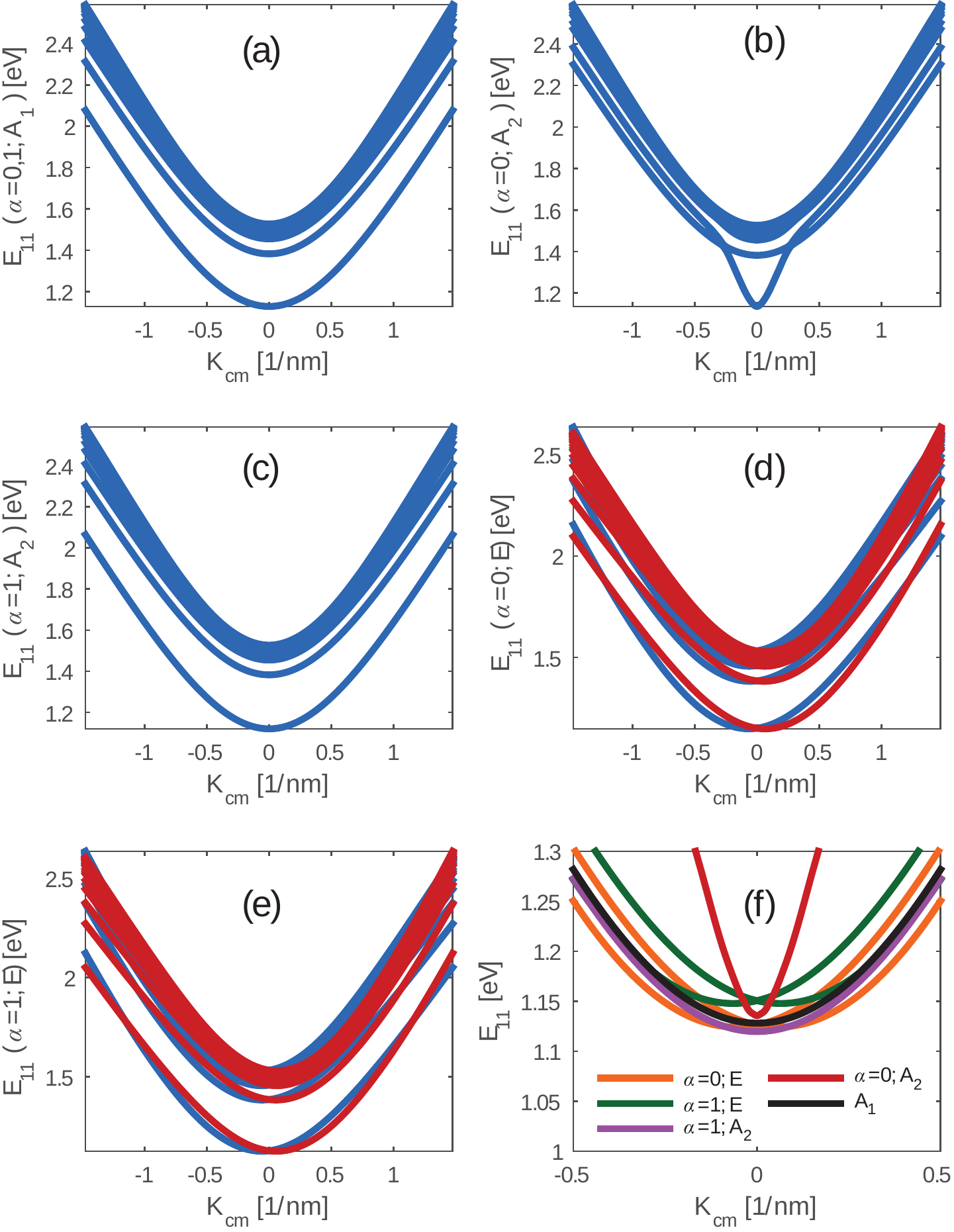}\\
  \caption{Energy dispersion of (a) \(A_1\) singlet and triplet exciton, (b) \(A_2\) singlet, (c) \(A_2\) triplet exciton, (d) \(E\) type singlet excitons with positive (blue) and negative (red) circumferential momentum, and (e) \(E\) type triplet excitons with positive (blue) and negative (red) circumferential momentum for \(S_{11}\) transition in (7,5) carbon nanotube. Panel (f) shows a comparison of lowest-subband exciton dispersions for various exciton types. $\alpha$ denotes the exciton spin, so $\alpha=0$ refers to singlet and $\alpha=1$ to triplet excitons.}
  \label{fig:ExcitonEnergy}
\end{figure}

We can calculate the exciton energies for different values of center-of-mass momentum, which yields the exciton dispersion curves. Figure \ref{fig:ExcitonEnergy} shows the energy dispersions for singlet and triplet excitons, with various symmetries and center-of-mass momentum. Here, it is assumed that the exciton is a result of an \(S_{11}\) transition (an electron is excited from the highest valence band to the lowest conduction band).

\section{Resonance energy transfer: direct interaction} \label{sec:resonance_energy_transfer_rate}
\begin{figure}[!tbp]
  \centering
  \includegraphics[width=0.8\linewidth]{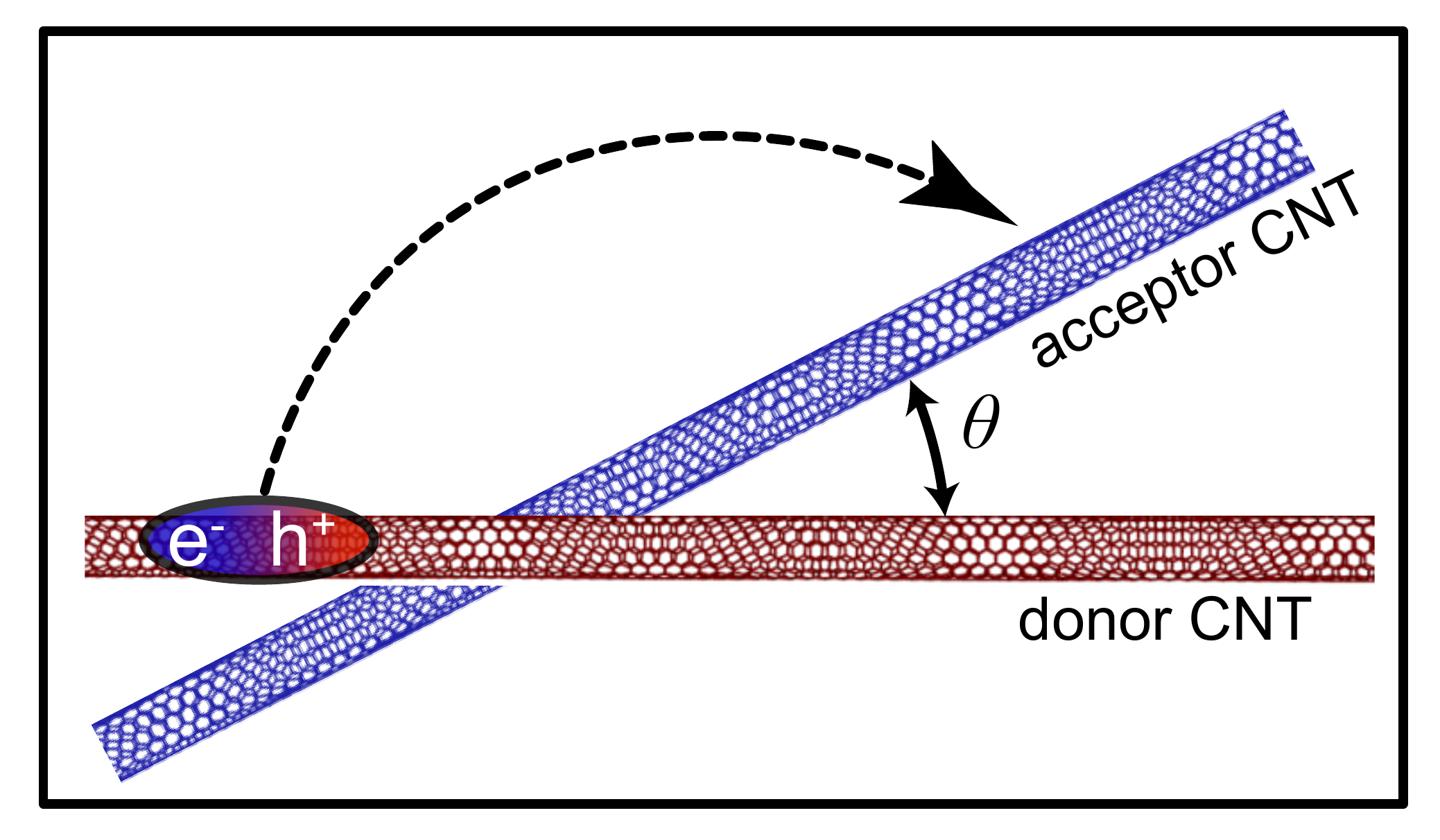}\\
  \caption{Geometry of donor (red) and acceptor (blue) carbon nanotubes.}
  \label{fig:exciton_transfer_geometry}
\end{figure}

In this section, we calculate the Coulomb interaction matrix element between excitonic states of two CNTs and the exciton transfer rate due to this interaction (Fig. \ref{fig:exciton_transfer_geometry}).

\begin{figure}[!tbp]
  \centering
  \includegraphics[width=0.8\linewidth]{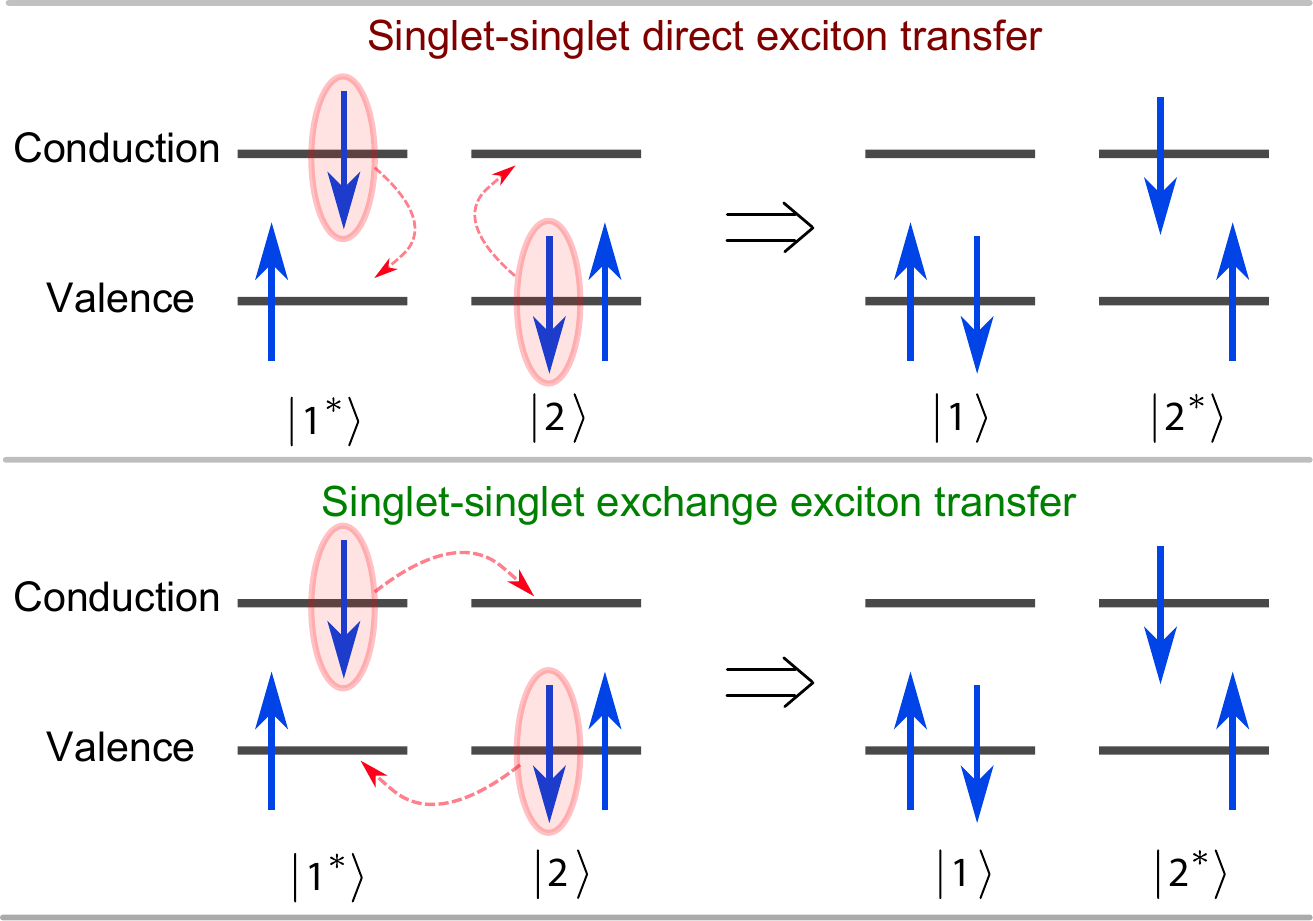}\\
  \caption{Schematic of the exciton transfer processes mediated by direct (top) and exchange (bottom) Coulomb interaction.}
  \label{fig:direct_exchange_transfer_schematic}
\end{figure}

The initial and final states of the system are assumed to be those of two noninteracting systems. Here, we assume that the first CNT is the donor CNT which is initially excited, and as a result of exciton scattering the excitation is transferred to the second, acceptor CNT. Therefore, the exciton initial and final states in this two-CNT system can be denoted by \(|I\rangle=|1^*\rangle \otimes |2\rangle\) and \(|F\rangle=|1\rangle \otimes |2^*\rangle\), respectively, where the asterisk denotes an excited state of a given tube. The scattering (transfer) process is possible through direct and exchange Coulomb interactions  between electrons of donor and acceptor CNTs \cite{Scholes2003long, May2008charge}. Figure \ref{fig:direct_exchange_transfer_schematic} shows a schematic of the direct and exchange pathways.

The exchange interaction is important for the CNT aggregates with considerable orbital overlaps between the donor and acceptor systems \cite{Scholes2003long, May2008charge}. When the separation between the donor and acceptor CNTs are larger than the spatial extent of the \(p_z\) orbitals ($<1\, \mathrm{\AA}$), the direct interaction is dominant. Here, we assume that the donor and acceptor CNTs are not touching (wall-to-wall separation $\approx 2\, \mathrm{\AA}$), hence we only consider the direct Coulomb interaction. The matrix element of the direct interaction is
\begin{equation} \label{eq:coulomb_matrix_element}
    \begin{split}
        \mathcal{M}_\text{d} = & \langle s_1,\bm{K}_1 ; \text{GS}_2 | \; \hat{\mathcal{H}}_\text{d} \; |\text{GS}_1 ; s_2,\bm{K}_2 \rangle \\
        = & \sum_{\bm{k}_{r_1}}\sum_{\bm{k}_{r_2}} A^*_{s_1}(\bm{K}_1,\bm{k}_{r_1}) A_{s_2}(\bm{K}_2,\bm{k}_{r_2}) \langle \bm{k}_{v_2}, \bm{k}_{c_1} |v(\bm{r}-\bm{r}')|\bm{k}_{c_2}, \bm{k}_{v_1} \rangle,
    \end{split}
\end{equation}

\noindent where we have \(\bm{k}_{c_1}=\bm{k}_{r_1}+\bm{K}_1\), \(\bm{k}_{v_1}=\bm{k}_{r_1}-\bm{K}_1\), \(\bm{k}_{c_2}=\bm{k}_{r_2}+\bm{K}_2\), and \(\bm{k}_{v_2}=\bm{k}_{r_2}-\bm{K}_2.\) The Coulomb interaction potential between the electrons is
\begin{equation}
    v(|\bm{r}-\bm{r}'|) = \frac{e^2}{4\pi\epsilon |\bm{r}-\bm{r}'|}.
\end{equation}

\noindent Here, $\epsilon=\epsilon_0\bar{\kappa}$ is the average value of the absolute dielectric permittivity of the generally nonuniform medium between the tubes; $\bar{\kappa}$ is the medium's average dielectric constant.

\noindent Now, we calculate the overlap integral over the free-particle tight-binding wave functions

\begin{equation}
    \begin{split}
        \mathcal{I}_{\text{TB}} = & \langle \bm{k}_{v_2}, \bm{k}_{c_1} |v(\bm{r}-\bm{r}')|\bm{k}_{c_2}, \bm{k}_{v_1} \rangle = \int \diff \bm{r}\, \diff \bm{r}' \psi_{c\bm{k}_{c_1}}^*(\bm{r}) \psi_{v\bm{k}_{v_1}}(\bm{r}) v(|\bm{r}-\bm{r}'|) \psi_{c\bm{k}_{c_2}}(\bm{r}') \psi_{v\bm{k}_{v_2}}^*(\bm{r}') \\
        = & \frac{1}{N_{u_1} N_{u_2}} \sum_{b_1,b_2}\sum_{b_3,b_4} C_{cb_1}^*(\bm{k}_{c_1})C_{vb_2}(\bm{k}_{v_1}) C_{cb_3}(\bm{k}_{c_2})C_{vb_4}^*(\bm{k}_{v_2}) \times \cdots \\
        & \hspace{0cm}\cdots \sum_{u_1,u_2}\sum_{u_3,u_4} e^{i(-\bm{k}_{c_1}.\bm{R}_{u_1b_1}+\bm{k}_{v_1}.\bm{R}_{u_2b_2}+\bm{k}_{c_2}.\bm{R}_{u_3b_3}-\bm{k}_{v_2}.\bm{R}_{u_4b_4})} \times \cdots \\
        & \hspace{0 cm} \cdots \bigg( \int \diff \bm{r}\, \diff \bm{r}' \phi^*(\bm{r}-\bm{R}_{u_1b_1}) \phi(\bm{r}-\bm{R}_{u_2b_2}) v(|\bm{r}-\bm{r}'|) \phi^*(\bm{r}'-\bm{R}_{u_3b_3}) \phi(\bm{r}'-\bm{R}_{u_4b_4}) \bigg).
    \end{split}
\end{equation}

\noindent Now, we assume that the last integral is important only when \(u_1b_1=u_2b_2=ub\) and \(u_3b_3=u_4b_4=u'b'\). Then the overlap integral becomes

\begin{equation}
    \begin{split}
        & \mathcal{I}_{\text{TB}} = \frac{1}{N_{u_1} N_{u_2}} \sum_{b,b'}C_{cb}^*(\bm{k}_{c_1})C_{vb}(\bm{k}_{v_1}) C_{cb'}(\bm{k}_{c_2})C_{vb'}^*(\bm{k}_{v_2}) \times \cdots \\
        & \hspace{0.5 cm} \cdots \sum_{u,u'} e^{i\left[(\bm{k}_{v_1}-\bm{k}_{c_1}).\bm{R}_{ub}+(\bm{k}_{c_2}-\bm{k}_{v_2}).\bm{R}_{u'b'}\right]}  \!\!\! \int \diff \bm{r}\, \diff \bm{r}' |\phi(\bm{r}-\bm{R}_{ub})|^2 v(|\bm{r}-\bm{r}'|) |\phi(\bm{r}'-\bm{R}_{u'b'})|^2.
    \end{split}
\end{equation}

\noindent Considering the small size of atomic orbitals with respect to the separation of the atoms in the two system, we can use a transition monopole approximation (TMA) \cite{Chang77,Wong2009ideal}

\begin{equation}
    \int \diff \bm{r}\, \diff \bm{r}' |\phi(\bm{r}-\bm{R}_{ub})|^2 v(|\bm{r}-\bm{r}'|) |\phi(\bm{r}'-\bm{R}_{u'b'})|^2\approx
    \frac{e^2}{4\pi\epsilon|\bm{R}_{ub}-\bm{R}_{u'b'}|}.
\end{equation}

\noindent Therefore, we have

\begin{equation}
    \begin{split}
    & \mathcal{I}_{\text{TB}} = \frac{e^2}{4\pi\epsilon N_{u_1} N_{u_2}} \sum_{b,b'}C_{cb}^*(\bm{k}_{c_1})C_{vb}(\bm{k}_{v_1}) C_{cb'}(\bm{k}_{c_2})C_{vb'}^*(\bm{k}_{v_2}) \times \cdots \\
    & \hspace{5cm} \cdots \sum_{u,u'} e^{i(-2\bm{K}_1.\bm{R}_{ub}+2\bm{K}_2.\bm{R}_{u'b'})}
    \frac{1}{|\bm{R}_{ub}-\bm{R}_{u'b'}|}.
    \end{split}
\end{equation}

\noindent Taking the wall-to-wall separation between the tubes large enough that relative positions of basis atoms in donor and acceptor CNTs are not important in calculating the exciton transfer rate, we have

\begin{equation}
    \frac{1}{|\bm{R}_{ub}-\bm{R}_{u'b'}|} \approx \frac{1}{|\bm{R}_{u}-\bm{R}_{u'}|}.
\end{equation}

\noindent Therefore, the matrix element is

\begin{equation} \label{eq:InteractionOverlapIntegral}
    \begin{split}
        & \mathcal{I}_{\text{TB}} = \frac{e^2}{4\pi\epsilon N_{u_1} N_{u_2}} J(\bm{K}_1,\bm{K}_2) \sum_{b,b'}C_{cb}^*(\bm{k}_{c_1}) C_{vb}(\bm{k}_{v_1}) C_{cb'}(\bm{k}_{c_2})C_{vb'}^*(\bm{k}_{v_2}) e^{i(-2\bm{K}_1.\bm{d}_{b}+2\bm{K}_2.\bm{d}_{b'})} \\
    \end{split}
\end{equation}

\noindent where we have used \(\bm{R}_{ub} = \bm{R}_u + \bm{d}_b\). $J(\bm
{K}_1,\bm{K}_2)$ is \emph{the geometric part of the matrix element} which contains all the information about relative orientation and position of donor and acceptor CNTs
\begin{equation} \label{eq:geometric_part_of_matrix_element}
    J(\bm{K}_1,\bm{K}_2) =  \sum_{u,u'} e^{i(-2\bm{K}_1.\bm{R}_{u}+2\bm{K}_2.\bm{R}_{u'})} \frac{1}{|\bm{R}_{u}-\bm{R}_{u'}|}.
\end{equation}

\noindent Therefore, the matrix element for direct Coulomb interaction is
\begin{multline}
    \mathcal{M}_\text{d}= \frac{e^2 J(\bm{K}_1,\bm{K}_2)}{4\pi\epsilon N_{u_1} N_{u_2}} \sum_{\bm{k}_{r_1}}\sum_{\bm{k}_{r_2}} A^*_{s_1}(\bm{K}_1,\bm{k}_{r_1}) A_{s_2}(\bm{K}_2,\bm{k}_{r_2}) \\
    \times \sum_{b,b'} \Bigg( C_{cb}^*(\bm{k}_{c_1})C_{vb}(\bm{k}_{v_1}) C_{cb'}(\bm{k}_{c_2})C_{vb'}^*(\bm{k}_{v_2}) \times \cdots \\
    \cdots e^{i(-2\bm{K}_1.\bm{d}_{b}+2\bm{K}_2.\bm{d}_{b'})} \Bigg) .
\end{multline}
{Note that, based on the normalization of the exciton wave function we have \(A_{s}(\bm{K},\bm{k}_r) \propto \frac{1}{\sqrt{N_u}}\), whereas the number of terms in the summation over \(\bm{k}_r\) increases linearly with \(N_u\). Therefore, we introduce the \(k\)-space part of the matrix element, which is independent of the length of donor and acceptor CNTs}
\begin{equation} \label{eq:coulomb_matrix_element_derived}
\begin{split}
    & Q(\bm{K}_1,\bm{K}_2) =
    \frac{e^2}{4\pi\epsilon \sqrt{N_{u_1} N_{u_2}}} \sum_{\bm{k}_{r_1}}\sum_{\bm{k}_{r_2}} A^*_{s_1}(\bm{K}_1,\bm{k}_{r_1}) A_{s_2}(\bm{K}_2,\bm{k}_{r_2}) \times \cdots \\
   & \hspace{4cm} \cdots \sum_{b,b'}C_{cb}^*(\bm{k}_{c_1})C_{vb}(\bm{k}_{v_1}) C_{cb'}(\bm{k}_{c_2})C_{vb'}^*(\bm{k}_{v_2}) e^{i(-2\bm{K}_1.\bm{d}_{b}+2\bm{K}_2.\bm{d}_{b'})}.
\end{split}
\end{equation}
\noindent The direct interaction matrix element becomes
\begin{equation}
    \mathcal{M}_\text{d} = \frac{1}{\sqrt{N_{u_1} N_{u_2}}} J(\bm{K}_1,\bm{K}_2) \times Q(\bm{K}_1,\bm{K}_2).
\end{equation}

\noindent Next, we calculate the exciton scattering rate from donor to acceptor CNTs presented with indices 1 and 2, respectively. {If the exciton-phonon and exciton-impurity interaction is stronger than the Coulomb coupling between CNTs, we can assume that the excitons in each CNT have an equilibrium thermal distribution because they are effectively thermalized within each CNT on a much shorter timescale than the intertube Coulomb-mediated transfer process.} Therefore, the exciton transfer rate is an average of the exciton scattering rate with the equilibrium thermal distribution. We use Fermi's golden rule to calculate the scattering rate due to direct interaction between CNTs

\begin{equation} \label{eq:fermi_transfer_rate}
    \Gamma_{12} = \frac{2\pi}{\hbar}\sum_{s_1,s_2}\sum_{\bm{K}_1,\bm{K}_2} \frac{e^{-\beta \Omega_{s_1}}}{\mathcal{Z}} |\mathcal{M}_\text{d} |^2 \delta(\Omega_{s_1}-\Omega_{s_2}),
\end{equation}

\noindent where \(\mathcal{Z}\) is the partition function
\begin{equation}
    \mathcal{Z}=\mbox{tr}\{e^{-\mathcal{H}/k_B T} \} = \sum_{s_1}\sum_{\bm{K}_1} e^{-\beta \Omega_{s_1}}.
\end{equation}

\noindent Assuming the tubes are long enough that we can convert the summation over \(\bm{K}_2\) into integration, we get

\begin{equation}
    \begin{aligned}
        \Gamma_{12} = & \frac{2\pi}{\hbar}\frac{1}{\Delta K_2} \sum_{s_1,s_2}\sum_{\bm{K}_1}\sum_{\mathscr{M}_2} \int dK_2 \frac{e^{-\beta \Omega_{s_1}}}{\mathcal{Z}} \left|\mathcal{M}_\text{d} \right|^2 \delta(\Omega_{s_1}-\Omega_{s_2}) \\
               = & \frac{2\pi}{\hbar}\frac{1}{\Delta K_2} \sum_{s_1,s_2} \sum_{\bm{K}_1} \sum_{\mathscr{M}_2} \frac{e^{-\beta \Omega_{s_1}}}{\mathcal{Z}} \left|\mathcal{M}_\text{d} \right|^2 \left( \frac{dK_2}{d\Omega_{s_2}}\right)_{\Omega_{s_1}}\\
               = & \frac{2\pi}{\hbar}\frac{1}{N_{u_1} N_{u_2} \Delta K_2} \sum_{s_1,s_2} \sum_{\bm{K}_1} \sum_{\mathscr{M}_2} \frac{e^{-\beta \Omega_{s_1}}}{\mathcal{Z}} \left|J(\bm{K}_1,\bm{K}_2) \times Q(\bm{K}_1,\bm{K}_2) \right|^2 \left( \frac{dK_2}{d\Omega_{s_2}}\right)_{\Omega_{s_1}}.
    \end{aligned}
\end{equation}
In order to calculate the transfer rate for a limited length CNT, it is better to write this equations in terms of CNT length, therefore, we use the following relations
\begin{equation}
    N_{u_1} = \frac{2 \pi r_1 L_1}{A_u} \quad , \quad N_{u_2} = \frac{2 \pi r_2 L_2}{ A_u} \quad , \quad \Delta K_2 = \frac{2\pi}{L_2}.
\end{equation}
Here, \(A_u\) is the area of the graphene unit cell. \(r_1\) and \(r_2\) are the radii of donor and acceptor CNTs, respectively. Therefore, we get
\begin{subequations}
    \begin{equation} \label{eq:exciton_transfer_rate}
        \Gamma_{12} =  \frac{1}{\hbar L_1} \left(\frac{e^2 A_u^2}{16\pi^3 r_1 r_2}\right)^2 \sum_{s_1,s_2} \sum_{\bm{K}_1} \sum_{\mathscr{M}_2} \frac{e^{-\beta \Omega_{s_1}}}{\mathcal{Z}} \left|J(\bm{K}_1,\bm{K}_2) \times \tilde{Q}(\bm{K}_1,\bm{K}_2) \right|^2 \left( \frac{dK_2}{d\Omega_{s_2}}\right)_{\Omega_{s_1}}\, ,
    \end{equation}
    where we have defined the renormalized \(k\)-part of the matrix element as
    \begin{equation}
    \begin{split}
        & \tilde{Q}(\bm{K}_1,\bm{K}_2) = \frac{1}{4\pi\epsilon \sqrt{L_1 L_2}} \sum_{\bm{k}_{r_1}}\sum_{\bm{k}_{r_2}} A^*_{s_1}(\bm{K}_1,\bm{k}_{r_1}) A_{s_2}(\bm{K}_2,\bm{k}_{r_2}) \times \cdots \\
        & \hspace{4cm} \cdots \sum_{b,b'}C_{cb}^*(\bm{k}_{c_1})C_{vb}(\bm{k}_{v_1}) C_{cb'}(\bm{k}_{c_2})C_{vb'}^*(\bm{k}_{v_2}) e^{i(-2\bm{K}_1.\bm{d}_{b}+2\bm{K}_2.\bm{d}_{b'})}\, .
    \end{split}
    \end{equation}
\end{subequations}

\section{Results and discussion} \label{sec:results_and_discussion}
\begin{figure}[!h]
  \centering
  \includegraphics[width=\linewidth]{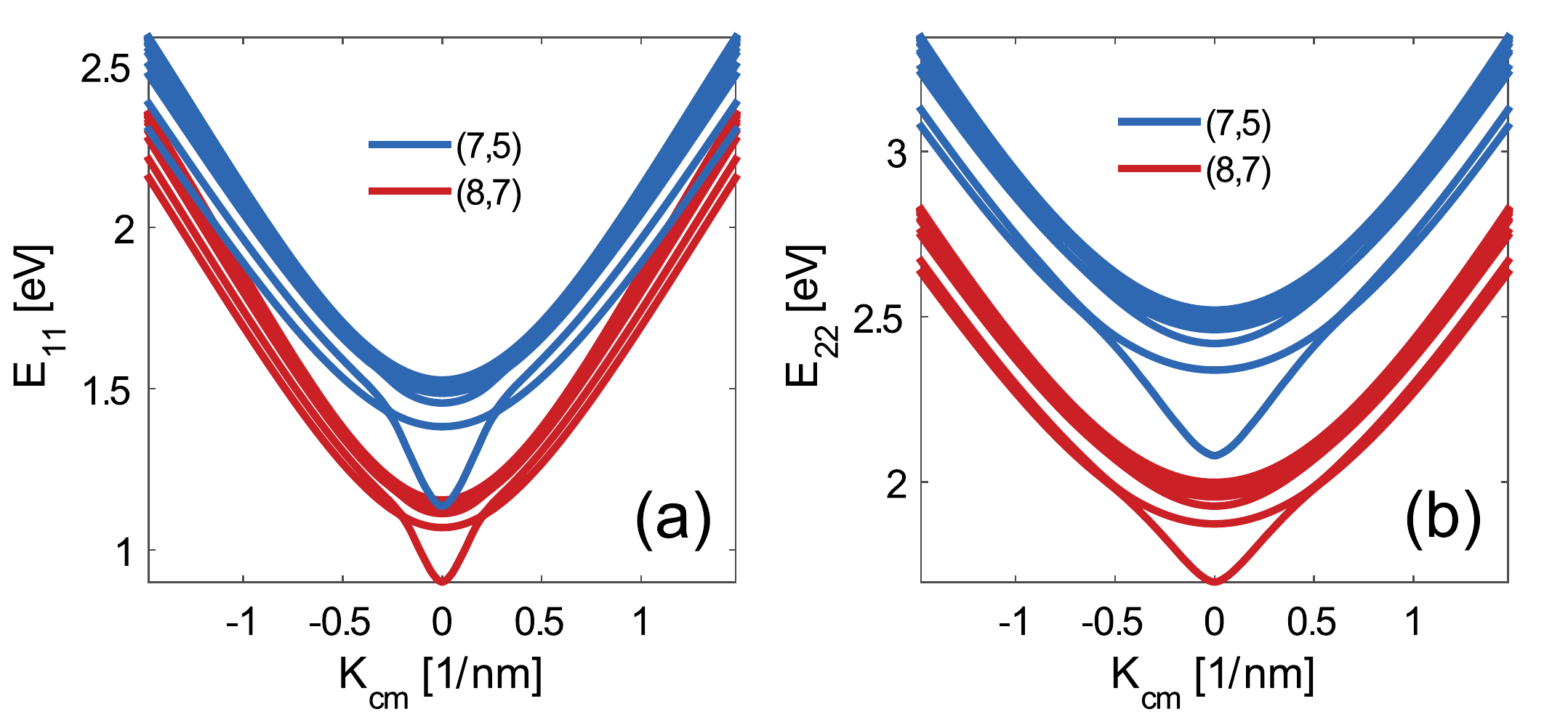}\\
  \caption{Comparison of $S_{11}$ (a) and $S_{22}$ (b)  excitonic energy dispersions in (7,5) and (8,7) CNTs.}
  \label{fig:exciton_dispersion_comparison_75_87}
\end{figure}

In this section, we calculate the exciton transfer rates between four different tube chiralities: (7,5), (7,6), (8,6), and (8,7). The energies of the lowest bright excitonic states in these CNTs are shown in Table \ref{tab:tab1}. Figure \ref{fig:exciton_dispersion_comparison_75_87} shows a comparison of the dispersions of optically active \(S_{11}\) and \(S_{22}\) excitonic states in (7,5) and (8,7) CNTs. We have taken the separation between the centers of donor and acceptor CNTs to be 1.2 nm, which provides enough wall-to-wall distance ($\approx 2\,\mathrm{\AA}$) between the CNTs under consideration that the exchange Coulomb  interaction is negligible.

In this section, we calculate the exciton transfer rates across different combinations of transition subbands (i.e., $S_{ii} \rightarrow S_{jj} $). Moreover, we calculate the exciton transfer rate between optically bright and optically dark excitonic states. We report on the dependence of the exciton transfer rate on the angle between the donor and acceptor tubes. Next, we study the effect of exciton confinement on the exciton transfer rate. Furthermore, we show that the exciton thermalization among both dark and bright states can reduce the exciton transfer rate by an order of magnitude. Also, we study the exciton-transfer-rate variation with varying electrostatic screening due to inhomogeneities in the surrounding medium.

\begin{table}[!tbp]
    \begin{tabular}{ l c r}
        \hline
        Chirality & \(S_{11}\) energy [eV] & \(S_{22}\) energy [eV] \\
        \hline
        \((7,5)\) & \(1.136 \) & \(2.08\) \\
        \((7,6)\) & \(1.053 \) & \(1.974\) \\
        \((8,6)\) & \(1.014 \) & \(1.849\) \\
        \((8,7)\) & \(0.9004 \) & \(1.697\) \\
        \hline
    \end{tabular}
    \caption{Energy of lowest bright excitonic states for selected SWNTs.}\label{tab:tab1}
\end{table}

\subsection{Interband and intraband exciton transfer rates}
\begin{figure}[!tbp]
  \centering
  \includegraphics[width=\textwidth]{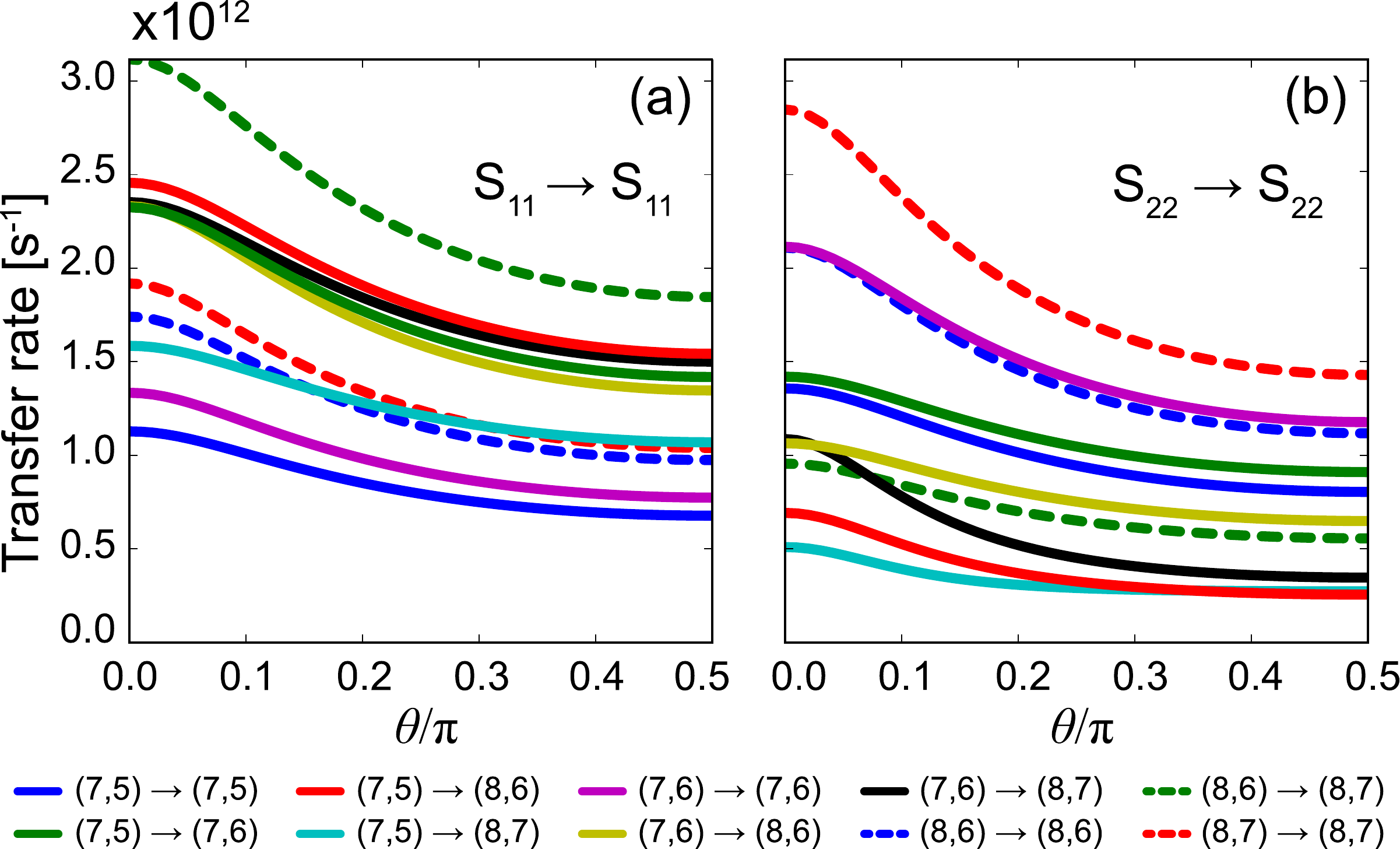}
  \caption{$A_2$ exciton transfer rate versus relative angle between donor and acceptor CNTs.  (a) $S_{11}\rightarrow S_{11}$  and (b) $S_{22}\rightarrow S_{22}$.}
  \label{fig:transfer_rate_versus_angle_A2_to_A2_S1_to_S1_and_S2_to_S2}
\end{figure}

First, we calculate the exciton transfer rate as a function of the relative angle between donor and acceptor CNTs (angle \(\theta\) in Fig. \ref{fig:exciton_transfer_geometry}). Due to the radiative nature of the direct resonance energy transfer (Figure \ref{fig:direct_exchange_transfer_schematic}), we expect the exciton transfer between bright excitonic states to be the dominant transfer process.  Figure \ref{fig:transfer_rate_versus_angle_A2_to_A2_S1_to_S1_and_S2_to_S2} shows the transfer rate of bright excitons (\(A_2\)) from donor CNTs with larger band gaps to acceptor CNTs with smaller band gaps (downhill transfer). The uphill exciton transfer process is usually a couple of orders of magnitude smaller,  because in this case the excitonic states in the donor tube that can resonate with the acceptor-tube states have low exciton population. In Fig. \ref{fig:transfer_rate_versus_angle_A2_to_A2_S1_to_S1_and_S2_to_S2}, the excitons belong to the same transition subbands in donor and acceptor CNTs (intraband exciton transfer). We assume excitons are confined inside a \(10 \text{ nm}\) long quantum well, similar to the work by Wong \textit{et al.} \cite{Wong2009ideal} We observe a relatively small dependence of the ET rate on the relative angle between CNTs. This is in contrast with the prediction by Wong \textit{et al.} \cite{Wong2009ideal}, who calculated that the transfer rate would drop to zero when the donor and acceptor tubes are perpendicular. This discrepancy stems from the method employed by Wong \emph{et al.}: they used a formulation that assumes that the Coulomb interaction matrix element, Eq. (\ref{eq:coulomb_matrix_element}), is almost constant between excitonic states with different energies. However, our detailed derivation shows that one needs to calculate the interaction matrix element for each pair of states in donor and acceptor CNTs [see Eq. (\ref{eq:coulomb_matrix_element_derived})]. In addition, our calculated rates is in excellent agreement with the study by Qian \emph{et al.}, who measured a lifetime of $\tau \approx 0.5 \text{ ps}$ for exciton transfer between two nonparallel CNTs \cite{Qian2008exciton}.

Moreover, the exciton transfer rate is slightly higher between \(S_{11}\) states than the transfer rate between \(S_{22}\) states. This lower transfer rate can be explained from the point of view that the direct resonance energy transfer is a process of simultaneous emission and absorption of a virtual photon by the donor and acceptor systems, respectively. The \(S_{22}\) excitonic states that are in resonance between the donor and acceptor tubes on average have a higher center-of-mass momentum, which yields a lower photon emission rate. \cite{Spataru2004excitonic, Malic2010excitonic} The lower rate of photon emission yields a lower exciton transfer rate between \(S_{22}\) states.

\begin{figure}[!tbp]
  \centering
  \includegraphics[width=\linewidth]{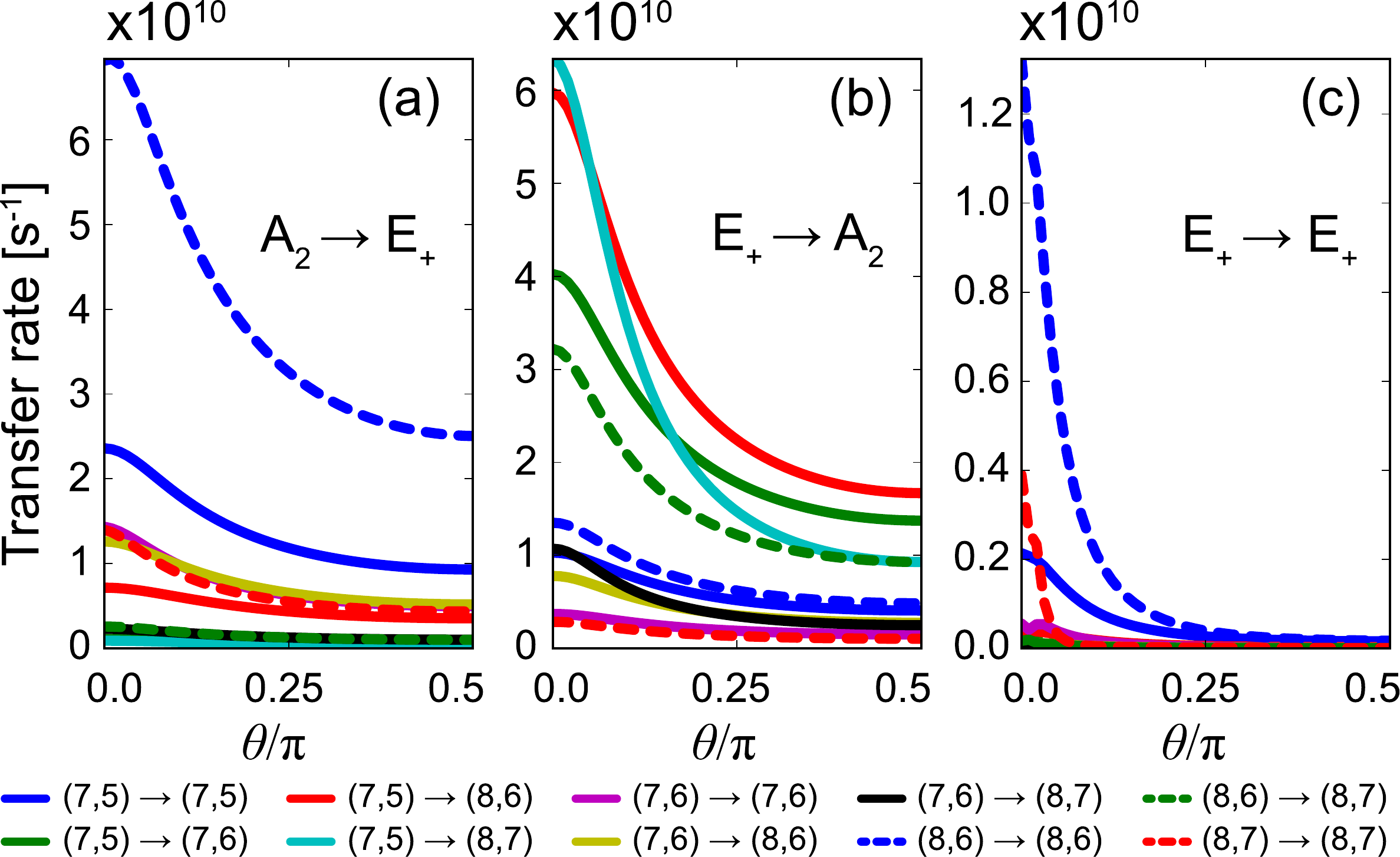}
  \caption{Transfer rate for processes involving dark $E_+$ excitons versus relative orientation of the donor and acceptor tubes and for different tube chiralities. (a) $A_2$ to $E_+$, (b) $E_+$ to $A_2$, and (c) $E_+$ to $E_+$.}
  \label{fig:transfer_rate_versus_angle_S1_to_S1_Ep_to_A2_and_A2_to_Ep_and_Ep_to_Ep}
\end{figure}

Next, we look at the contribution of dark excitonic states in the exciton transfer process. When the separation between the donor and acceptor molecules is larger than the size of each molecule, the traditional F\"orster theory is applied to calculate the exciton transfer rate. In this case, the transfer process depends on the overlap between the emission and absorption spectra of the donor and acceptor molecules, respectively. Therefore, the dark excitonic states do not contribute in transfer process. However, in the case of exciton transfer between neighboring CNTs, the F\"orster theory fails as the donor and acceptor molecules are relatively large \cite{Wong2009ideal}. Therefore, some dark excitonic states could contribute to the energy transfer process. Figure \ref{fig:transfer_rate_versus_angle_S1_to_S1_Ep_to_A2_and_A2_to_Ep_and_Ep_to_Ep} shows the downhill ET rate to or from dark \(E\)-type excitonic states among four CNT types. These transfer rates are about two orders of magnitude smaller than the transfer rates between bright excitonic states (Fig. \ref{fig:transfer_rate_versus_angle_A2_to_A2_S1_to_S1_and_S2_to_S2}), owing to the large angular momentum of \(E\)-type excitons. However, Postupna \emph{et al.} \cite{Postupna2014photoinduced} suggested that the phonon-assisted processes could facilitate efficient exciton transfer from/to dark excitonic states \cite{Postupna2014photoinduced}. Postupna \emph{et al.} used time-dependent density functional theory (TDDFT) in conjunction with molecular dynamic (MD) to study phonon-assisted exciton hopping. Due to numerical reasons, their study was limited to the ET process between two arrays of CNTs with a relative angle of $90^{\circ}$.

We should note that the dark \(A_1\) excitonic states still do not contribute to the exciton transfer process due to symmetry considerations that hold beyond the dipole approximation \cite{Jiang2007chirality}.

\begin{figure}[!tbp]
  \centering
  \includegraphics[width=\linewidth]{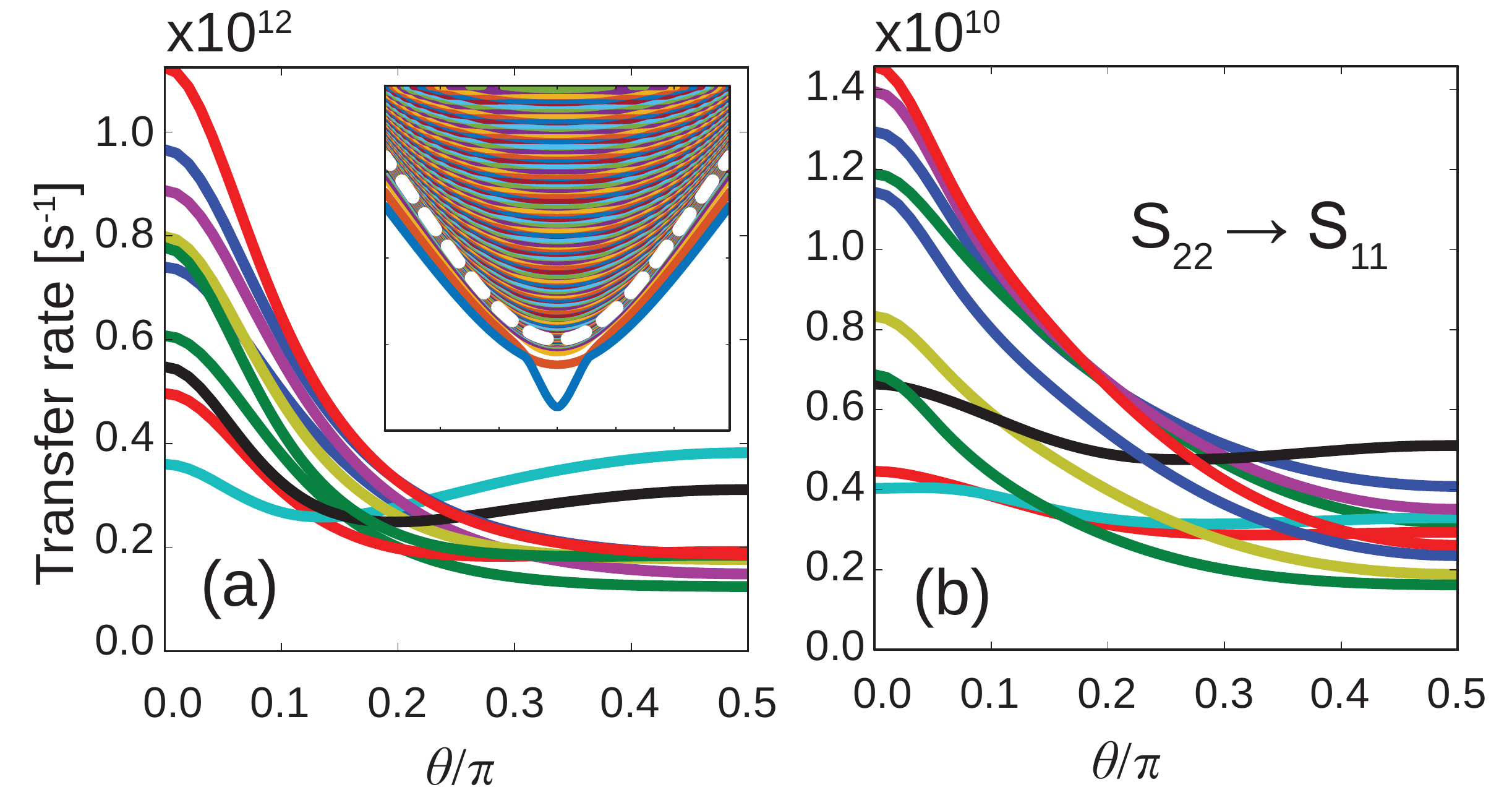}\\
  \caption{{(a) Transfer rate of bright $S_{22}$ excitonic states to bright $S_{11}$ excitonic states as a function of tube orientation when considering all excitonic states, both tightly bound and continuum. (Inset) Dispersions of bound and continuum excitonic states, separated by a dashed white line that denotes the lowest continuum level. (b) Transfer rate of bright $S_{22}$ excitonic states to bright $S_{11}$ excitonic states as a function of tube orientation when considering only bound excitonic states.}}
  \label{fig:transfer_rate_versus_angle_A2_to_A2_S2_to_S1}
\end{figure}

{Next, we look at the exciton transfer process from the \(S_{22}\) excitonic states in the donor tube to the \(S_{11}\) excitonic states in the acceptor tube (interband exciton transfer). As we can see in Fig. \ref{fig:transfer_rate_versus_angle_A2_to_A2_S2_to_S1}, the interband energy transfer process occurs almost as fast as the intraband exciton transfer process shown in Fig. \ref{fig:transfer_rate_versus_angle_A2_to_A2_S1_to_S1_and_S2_to_S2}. In order to better understand the role of different excitonic states in the excitation energy transfer process, we calculated the intraband ($S_{11} \rightarrow S_{11}$ and $S_{22}\rightarrow S_{22}$) and interband ($S_{22} \rightarrow S_{11}$) exciton transfer rates considering only the exciton transfer process between tightly bound excitonic states below the continuum level
\bibnote{The continuum level is the minimum energy level beyond which the excitonic state can essentially be considered as free electron and free hole states.}
(white dashed line in the inset to Fig. \ref{fig:transfer_rate_versus_angle_A2_to_A2_S2_to_S1}a). The calculated intraband transfer rates did not change significantly from the case that included both tightly bound excitonic states and the continuum states (in other words, they remain similar to those depicted in Fig. \ref{fig:transfer_rate_versus_angle_A2_to_A2_S1_to_S1_and_S2_to_S2}). However, the interband exciton transfer rates decreased by about two orders of magnitude when the continuum states were eliminated (compare Fig. \ref{fig:transfer_rate_versus_angle_A2_to_A2_S2_to_S1}a to Fig. \ref{fig:transfer_rate_versus_angle_A2_to_A2_S2_to_S1}b). We conclude that, although there are many transition states in the continuum region that resonate between the donor and acceptor states, they do not contribute to the intraband exciton transfer process. In contrast, most of the interband exciton transfer processes occur from tightly bound excitonic states to these continuum states.}

\subsection{Exciton confinement effect} \label{sec:exciton_confinement_effect}
{Owing to various forms of disorder (e.g., the nonuniformity in the dielectric properties of the surrounding environment and the presence of charged impurities in the CNT samples), an exciton can be confined in quantum wells along the CNT. In this section, we study the effect of quantum-well size on the exciton transfer rate. For relatively wide quantum wells, which yield excitonic states with small energy spacing, the matrix element of the Coulomb interaction and the calculated ET rate are affected only by the change in the geometric part of the matrix element [Eq. (\ref{eq:geometric_part_of_matrix_element})]. Therefore, we calculate the ET rate through Eq. (\ref{eq:exciton_transfer_rate}), where the spatial extent of the quantum well is used in calculating the geometric part of the matrix element [Eq. (\ref{eq:geometric_part_of_matrix_element})].}

\begin{figure}[!tbp]
  \centering
  \includegraphics[width=\linewidth]{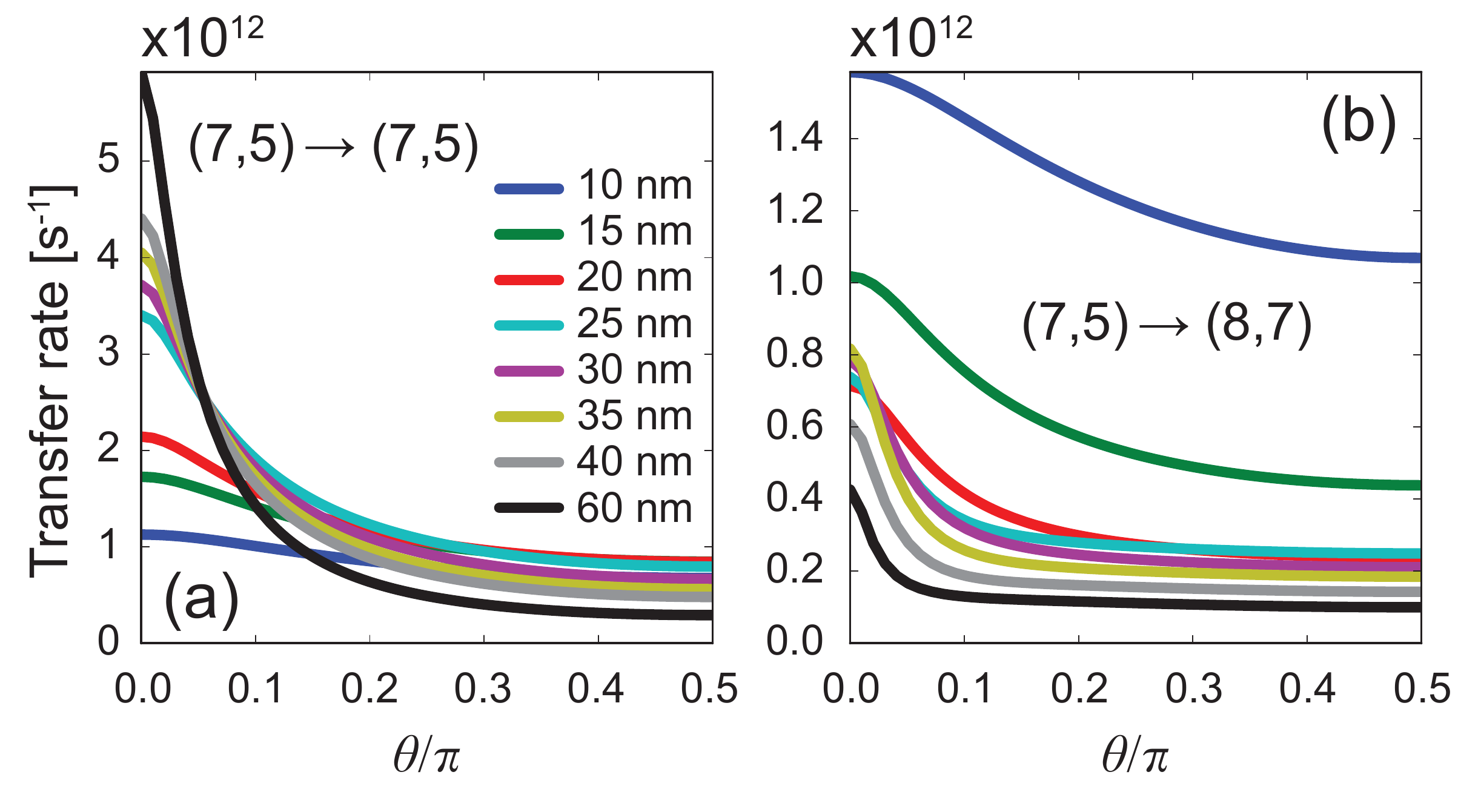}
  \caption{Exction transfer rate from bright excitonic states of a donor (7,5) SWNT to the bright excitonic states of (7,5) (panel a) and (8,7) (panel b) acceptor SWNTs. Different colors show the transfer rates of excitons with various quantum well sizes.}
  \label{fig:transfer_rate_versus_angle_various_length}
\end{figure}

Figure \ref{fig:transfer_rate_versus_angle_various_length}a (Figure \ref{fig:transfer_rate_versus_angle_various_length}b) shows the exciton transfer rate between bright excitonic states when the donor and acceptor SWNTs have similar (different) chiralities. It is assumed that the sizes of quantum wells in donor and acceptor SWNTs are the same.

\begin{figure}[!tbp]
  \centering
  \includegraphics[width=\linewidth]{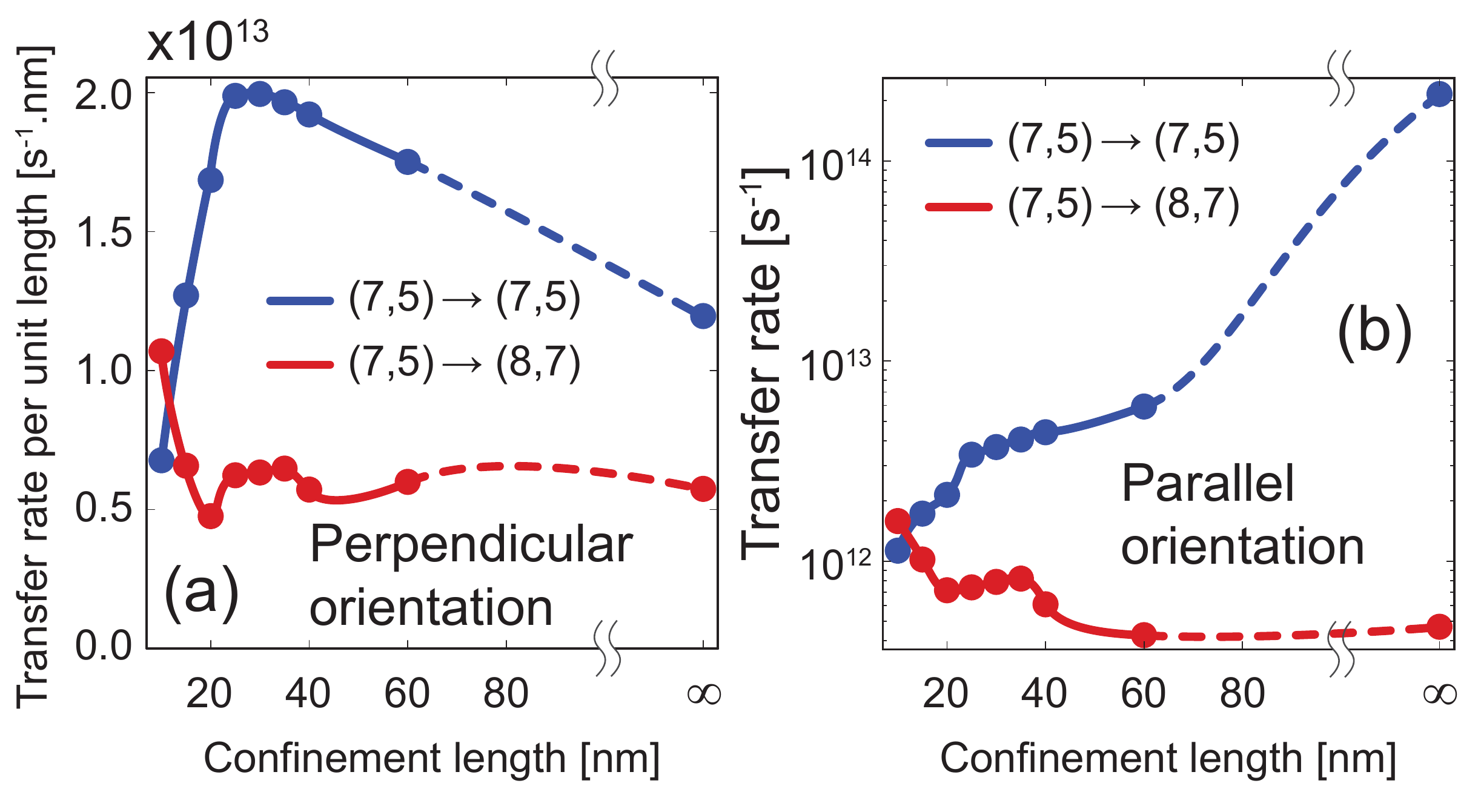}
  \caption{(a) The exciton transfer rate per unit length between perpendicular donor and acceptor SWNTs as a function of exciton confinement length. (b) The exciton transfer rate between parallel donor and acceptor SWNTs as a function of exciton confinement length.}
  \label{fig:transfer_rate_versus_cnt_length}
\end{figure}

When the CNTs are not parallel, the exciton transfer rate drops with increasing size of the quantum wells because the average spacing between the donor and acceptor systems increases. We can see this length dependence in Eq. (\ref{eq:exciton_transfer_rate}). However, in a CNT sample with a constant density of tubes, the number of available acceptor tubes is proportional to the length of the donor CNT. Therefore, we introduce the exciton transfer rate per unit length of the donor tube (Fig. \ref{fig:transfer_rate_versus_cnt_length}a). The exciton transfer rate per unit length changes up to a factor of three due to the variation of the geometric part of matrix element [Eq. (\ref{eq:exciton_transfer_rate})]. Nevertheless, the transfer rate stays relatively constant as we go to the limit of free-exciton transfer rates. The derivation of an analytical expression for the matrix element and the exciton transfer rate in the case of a free exciton in infinitely long donor and acceptor tubes is shown in Appendix  \ref{sec:derivation_of_free_exciton_transfer_rate}.

We observe a different behavior when the donor and acceptor tubes are parallel. This case is particularly important because in many samples the CNTs stick together and form CNT bundles. Unlike the case of nonparallel CNTs, we do not observe a drop in the exciton transfer rate with increasing size of quantum wells, as the average distance between the donor and acceptor states is almost independent of this size. We can predict this behavior based on the analytical expressions derived in Appendix \ref{sec:derivation_of_free_exciton_transfer_rate}.

Furthermore, based on the chirality of the donor and acceptor CNTs, the exciton transfer rates follow different trends as the size of quantum wells increases. For small confinement lengths, the exciton center-of-mass momenta in initial and final states are not important factors in determining the strength of Coulomb coupling; therefore, there are many states in the acceptor tube that can resonate with the donor-tube excitonic states. However, as the confinement length increases,  the contribution from the excitonic states that do not conserve momentum drops. In the limit of free-exciton transfer, only the states that conserve both the center-of-mass momentum and energy can transfer between the CNTs. Therefore, when the excitonic energy dispersions in the donor and acceptor tubes are dissimilar (e.g., due to different chiralities), a limited number of states contribute to the exciton transfer process. This results in a decrease of the exciton transfer rate. On the other hand, if the donor and acceptor CNTs have similar dispersion curves, there are many excitonic states that conserve both momentum and energy in donor and acceptor CNTs, which increases the exciton transfer rate by about two orders of magnitude (Fig. \ref{fig:transfer_rate_versus_cnt_length}).

These findings are in excellent agreements with measurements. L\"uer \emph{et al.} have measured ET rates between $S_{11}$ states within bundles of CNTs that exceed $10^{14}\;\text{s}^{-1}$  \cite{Luer2010ultrafast}. They also reported limited ET rate between $S_{22}$ states which is due to the same momentum matching considerations that we have discussed here. Recently, Mehlenbacher \emph{et al.} studied ET in samples where CNTs are wrapped in polymers and in samples with no residual surfactant.\cite{Mehlenbacher2013photoexcitaion, Mehlenbacher2015energy, Mehlenbacher2016ultrafast}. They found that, in the samples with no polymer wrapping, the ET rate between parallel CNTs is extremely fast, with $<60 \text{ fs}$ time scales. Two-dimensional anisotropy measurements showed much slower ET rates between nonparallel CNTs in these samples. On the other hand, Mehlenbacher \emph{et al.} found  picosecond time scales for ET rates in samples that CNTs are wrapped in polymers. In these samples, ET shows no preference between parallel and nonparallel relative orientation of donor and acceptor CNTs. Our calculations agree very well with these experimental findings: in pristine samples (thus no exciton confinement), we expect high transfer rates between same-chirality parallel tubes ($~10^{14}\;\text{s}^{-1}$) and much lower rates when the tubes are misoriented. In samples with polymer residue, excitons exhibit confinement,  which drastically reduces the rate of transfer between parallel tubes and results in isotropic ET rates of around $~10^{12}\;\text{s}^{-1}$.

\subsection{Effect of static screening by surrounding media}

\begin{figure}[!tbp]
  \centering
  \includegraphics[width=\linewidth]{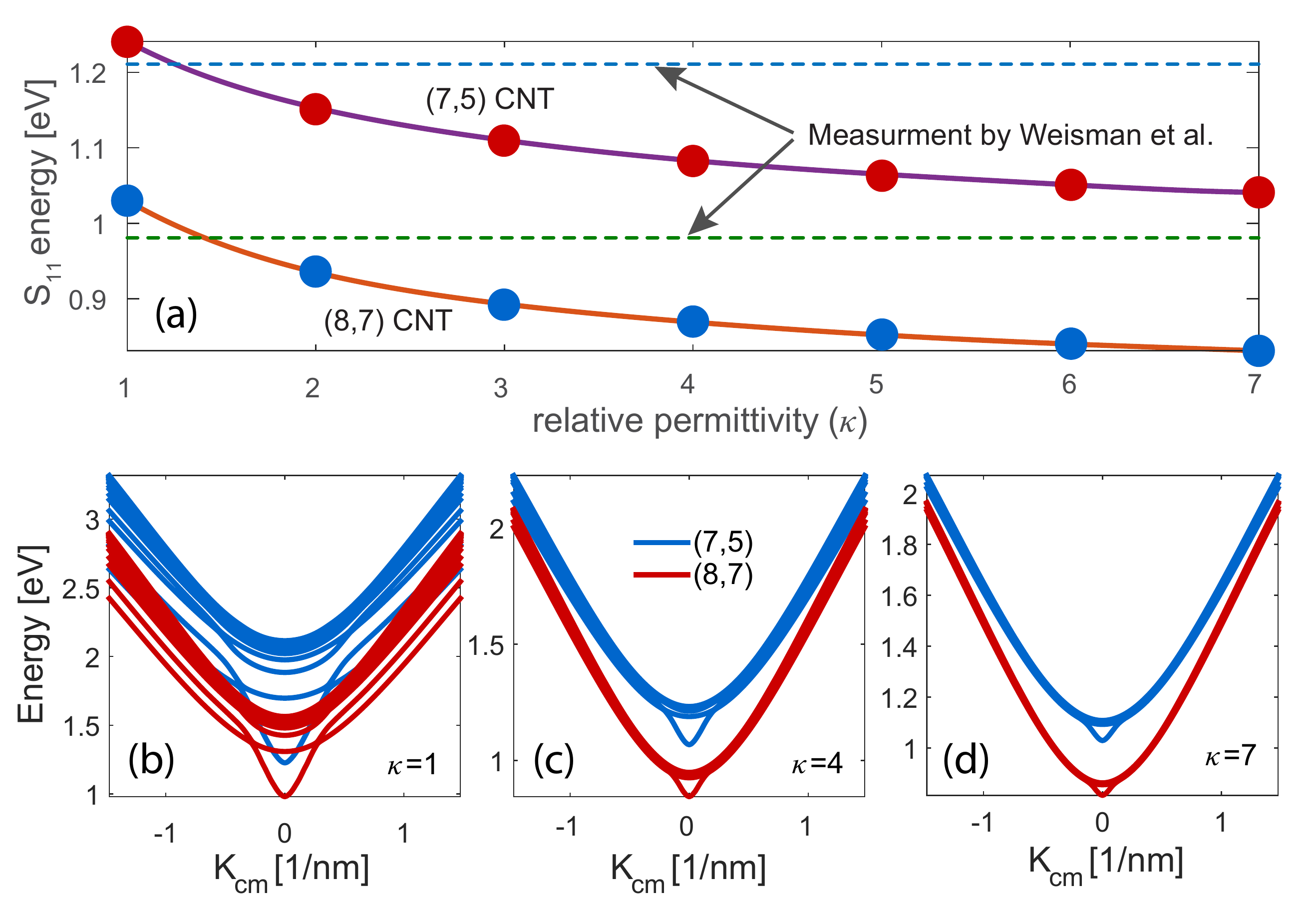}
  \caption{(a) Lowest transition energies in (7,5) and (8,7) SWNTs as a function of relative permittivity. (b) -- (d) Exciton dispersions with various relative dielectric permittivities of the environment.}
  \label{fig:transition_energy_vs_dielectric_constant}
\end{figure}

In this section, we study another effect of the CNT surrounding environment on the exciton transfer process. Like most quasi-one-dimensional nanostructures, the electronic and optical properties of CNTs are influenced by their surrounding media. One of these environmental effects is the screening of electrostatic electron-electron and electron-hole interactions inside a CNT. As we discussed before, the relative dielectric permittivity \({\kappa}\) in Eq. (\ref{eq:screened_coulomb_interaction}) accounts for the screening due to the surrounding medium and the core electrons in CNTs. As \(\kappa\) increases,  the self-energy due to repulsive electron-electron interaction and the binding energy due to the attractive electron-hole interaction decreases \cite{Araujo2009diameter}. As shown in Fig. \ref{fig:transition_energy_vs_dielectric_constant}a, the net effect is a decrease in exciton energy with increasing permittivity. In the limit of infinite permittivity, we retrieve the noninteracting electron results. Figure  \ref{fig:transition_energy_vs_dielectric_constant} show the energy dispersions for bright excitonic states assuming various values of $\kappa$. As expected, the binding energy and the number of tightly bound excitonic states decrease with increasing permittivity.

\begin{figure}[!tbp]
  \centering
  \includegraphics[width=\linewidth]{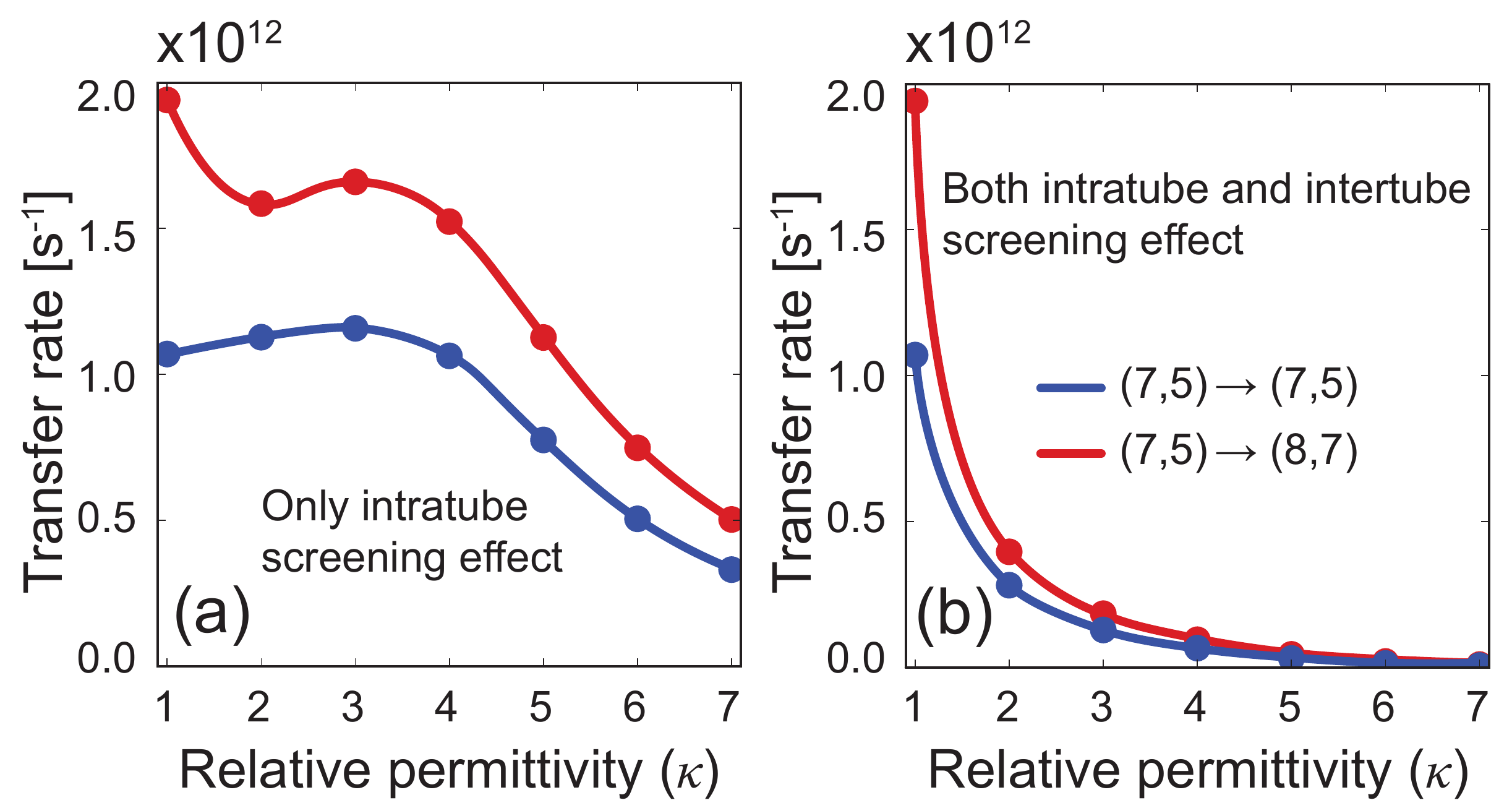}
  \caption{{Exciton transfer rate between (7,5) and (8,7) CNTs as a function of environment relative permittivity when (a) only the intratube screening effect is taken into account and (b) both intertube and intratube screening effects are taken into account.}}
  \label{fig:transfer_rate_versus_dielectric_function}
\end{figure}

{Screening of the Coulomb interactions via the surrounding medium  affects the ET rates in two ways: by changing the exciton wave function and dispersion energy (intratube screening effect) and by changing the electron-electron interaction between the donor and acceptor CNTs (intertube screening effect). The first effect stems from the change in the permittivity of the environment in the small area around the donor/acceptor CNT (we denote this permittivity by $\kappa$), whereas the second effect happens through changes in the average permittivity of the environment over long distances in the sample (we capture this effect through $\bar{\kappa}$). While $\bar{\kappa}$ and $\kappa$ are generally not the same, for simplicity here we assume that they are. Figure \ref{fig:transfer_rate_versus_dielectric_function}a shows the exciton transfer rate between bright excitonic states as a function of the environment relative permittivity when only the first effect is taken into account. The drop in the ET rate with increasing relative permittivity can be explained by looking at the photon absorption and emission rates. For higher permittivities, the excitonic states are more like free-electron and free-hole states than like bound excitons, and thus have lower photon absorption and emission rates \cite{Malic2010excitonic}. Consequently, the rate of ET, which is simultaneous photon emission and photon absorption by the two tubes, decreases. Figure \ref{fig:transfer_rate_versus_dielectric_function}b shows the exciton transfer rate between bright excitonic states as a function of Coulomb screening when accounting for both effects of screening. In this case, the drop in the ET rate with increasing permittivity is much more significant than in the previous case. This major drop is a result of a smaller perturbing Hamiltonian, which has a dependence of $\hat{\mathcal{H}}_{\text{d}} \propto \tfrac{1}{\kappa}$.}

\subsection{Interband exciton thermalization}
As we discussed earlier, the excitonic states in CNTs are classified as bright and dark states. The former exciton type is created via optical stimulation of ground-state electrons and the latter is usually populated through some second-order processes, such as Raman scattering or the scattering of bright excitons into dark excitons by phonons and impurities. So far, we have studied the intrinsic exciton transfer rate from either bright or dark excitonic states. However, if the exciton scattering between bright and dark states is fast enough (compared to the exciton transfer process), the excitons are thermalized among both bright and dark states and we have to consider both in the transfer process.

\begin{figure}[!tbp]
  \centering
  \includegraphics[width=1.0\linewidth]{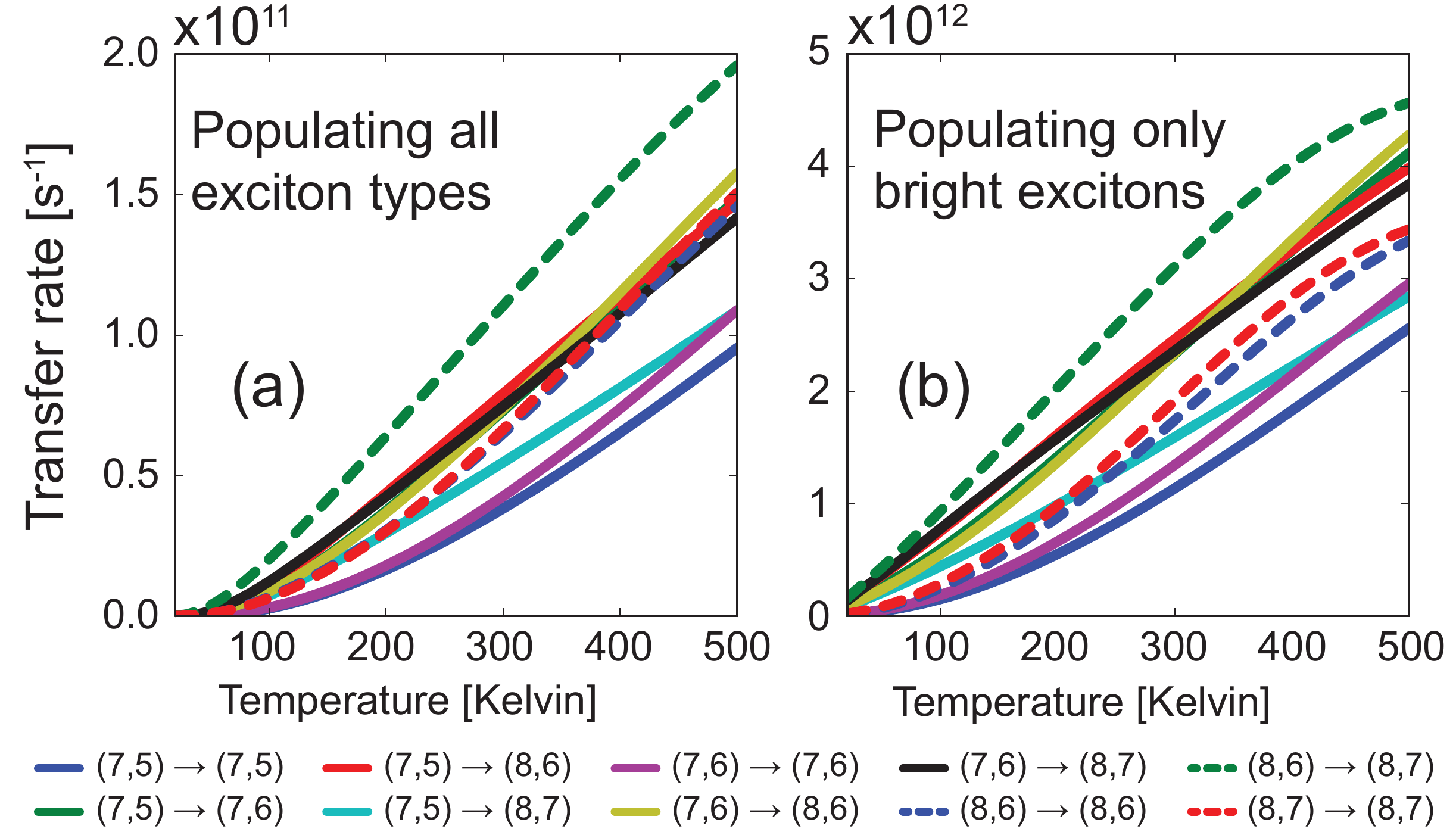}
  \caption{The exciton transfer rate of excitons between CNTs (a) when the excitons are thermalized between both bright and dark excitonic states and (b) when only the bright excitonic states are populated. The donor and acceptor CNTs are parallel and the excitons are assumed to be confined in a 10 nm wide quantum well.}
  \label{fig:transfer_rate_versus_temperature_confined_exciton}
\end{figure}

Figure \ref{fig:transfer_rate_versus_temperature_confined_exciton} shows the exciton transfer rate as a function of temperature for the full exciton thermalization among bright and dark excitonic states (Fig.  \ref{fig:transfer_rate_versus_temperature_confined_exciton}a) and among bright states only (Fig.  \ref{fig:transfer_rate_versus_temperature_confined_exciton}b). In the presence of exciton thermalization process between bright and dark excitonic states, we observe a twentyfold decrease in the exciton transfer rate. This is due to the presence of low-lying triplet states and the symmetric singlet states that do not transfer through the direct Coulomb interaction. As the temperature decreases to \(T=0\; \text{K}\), the excitons only populate these low-lying states and the transfer rate goes to zero.{In addition, we note that the intrisic transfer rate between bright excitonic decreases when temperature drops. This trend is contrary to the behavior of radiative exciton decay rate predicted by Perebeinos \emph{et al.} \cite{Perebeinos2005radiative}}

\section{Conclusion} \label{sec:summary}
{In summary, we calculated the exciton transfer rates between semiconducting CNTs of different chiralities and relative orientations. The exciton transfer rate is weakly dependent on the orientation of the tubes. This finding is in contradiction with previous theoretical studies, but in good agreement with experiments. The exciton transfer rate between bright excitonic states is about \(2  \times 10^{12} \text{ s}^{-1}\). The transfer rates between bright and dark states are at least two orders of magnitude smaller that the transfer rates between bright excitonic states. We also looked at the exciton transfer from \(S_{22}\) to \(S_{11}\) transition energies. We found that this type of exciton transfer process is as fast as transfer from $S_{11}$ to $S_{11}$ and from $S_{22}$ to $S_{22}$ states. This process is facilitated by coupling of tightly bound excitonic states in \(S_{22}\) to the continuum level states (equivalent of free-electron/free-hole states) in \(S_{11}\).

Furthermore, we studied the environmental effects on the exciton transfer rates. We calculated the exciton transfer rate for excitons confined to quantum wells with various sizes. When the quantum-well size increases, we observed a decrease in the transfer rate between nonparallel tubes. This is due to the increase in the average distance between donor and acceptor systems. By introducing a more relevant quantity (i.e., transfer rate per donor tube length), we showed that the exciton transfer rate is almost independent of the exciton confinement length. However, exciton transfer between parallel tubes follows a different trend: transfer between same-chirality donor and acceptor tubes is extremely sensitive to confinement and free excitons have the highest transfer rate (\(> 10^{14} \text{ s}^{-1}\)). The transfer rates between different-chirality tubes are not as sensitive to confinement and only  slightly increase with decreasing confinement length. These findings are a result of momentum conservation rules.

Moreover, we looked at the effect of Coulomb screening due to the surrounding media. The exciton transfer rate decreases with increasing screening. We also  showed that the exciton transfer rate increases with increasing temperature. This behavior is the opposite of what one would expect based on the emission and absorption spectra of donor and acceptor systems. Also, we showed that the exciton transfer rate drops by about one order of magnitude if excitons are thermalized between bright and dark excitonic states via extrinsic scattering sources (e.g., impurities and phonons).

We conclude that the wide range of ET-rate measurements, spanning two orders of magnitude, stems from variations in sample preparation and thus the degree of environmental disorder and homogeneity, as the ET rates in pristine samples and in the samples in which environmental disorder results in exciton confinement differ by two orders of magnitude.}

\begin{appendix}

\section{Derivation of the exciton transfer rate for free excitons} \label{sec:derivation_of_free_exciton_transfer_rate}

In this Appendix, we derive the transfer rate of free excitons between two very long CNTs of radii $r_1$ and $r_2$. We assume that the CNTs are far enough that, from the point of view of each CNT, the other one looks like a continuous  medium, so we can convert the last sum in Eq. (\ref{eq:geometric_part_of_matrix_element}) to an integral:
\begin{equation}
    \sum_{u,u'} e^{i(-2\bm{K}_1.\bm{R}_{ub}+2\bm{K}_2.\bm{R}_{u'b'})} \frac{1}{|\bm{R}_{ub}-\bm{R}_{u'b'}|} \\
    \approx \frac{1}{A_u^2} \times \int d^2\bm{r} \; d^2\bm{r}' \frac{e^{-2i\bm{K}_1.\bm{r} } \times e^{2i\bm{K}_2.\bm{r}'}} {|\bm{r}-\bm{r}'|}.
\end{equation}
Assuming that the CNTs are shifted by a center-to-center distance \(D\) along the \(z\)-axis and are misoriented by angle \(\theta\) in the \(xy\) plane, the position vectors are
\begin{equation}
    \begin{split}
        \bm{r} = & \; x \hat{\bm{x}} + r_1 \cos \phi \hat{\bm{y}} + r_1 \sin \phi \hat{\bm{z}},\\
        \bm{r}' = & \; (x'\cos \theta - r_2 \cos \phi' \sin \theta) \hat{\bm{x}} + (x' \sin \theta + r_2 \cos \phi' \cos \theta) \hat{\bm{y}} + (D + r_2 \sin \phi')\hat{\bm{z}}.
    \end{split}
\end{equation}

\noindent Therefore, the geometric part of the matrix element becomes

\begin{equation} \label{eq:ArbitraryAngleIntegral}
    \begin{split}
    &J_\theta(\bm{K}_1, \bm{K}_2) = \frac{r_1 r_2}{A_u^2} \int dx ~ dx' ~ d\phi ~ d\phi' e^{-2i(K_1 x+\mathscr{M}_1\phi)+2i(K_2x'+\mathscr{M}_2\phi')} \times \cdots \\
    &\hspace{1cm} \cdots \left[ (x'\cos \theta - r_2 \cos \phi' \sin \theta - x)^2 + \cdots \right.\\
    &\left.\hspace{2.2cm} \cdots (x' \sin \theta + r_2 \cos \phi' \cos \theta - r_1 \cos \phi )^2 + (D + r_2 \sin \phi' - r_1 \sin \phi)^2 \right]^{-\frac{1}{2}},
    \end{split}
\end{equation}

\noindent If the CNTs are infinitely long, the integrals over \(x\) and \(x'\) in the geometric part of matrix element can be calculated analytically \cite{Bateman1954tables}

\begin{equation}
    \begin{split}
        &J_\theta(\bm{K}_1,\bm{K}_2) = \frac{\pi r_1 r_2}{{A_u}^2} \int d\phi \, d\phi' \, \exp\big(2i(\mathscr{M}_2\phi'-\mathscr{M}_1\phi)\big) \times \cdots \\
        &\hspace{1cm} \cdots \exp\left(2i \frac{K_1(r_2 \cos\phi'-r_1 \cos\phi \cos\theta)+K_2(r_1 \cos\phi-r_2 \cos\phi' \cos\theta)}{\sin\theta} \right) \times \cdots \\
        &\hspace{5.2cm} \cdots \frac{\exp(-2 \frac{|D + r_2 \sin \phi' - r_1 \sin \phi|}{\sin\theta}\sqrt{ K^2_2 + K^2_1 -2K_1 K_2 \cos\theta})}{\sqrt{ K^2_2 + K^2_1 -2K_1 K_2 \cos\theta}}.
    \end{split}
\end{equation}

\noindent Using this relation in Eq. (\ref{eq:exciton_transfer_rate}), we can calculate the exciton transfer rate for free excitons. However, the transfer rate between two CNTs goes to zero due to infinitely long donor CNT. However, we can consider the transition rate from a donor CNT to an array of CNTs that are misoriented by angle \(\theta\). This is effectively equivalent to having a network of CNTs. The number of acceptor CNTs that the exciton can be transferred to from the initial donor CNT with length \(L_1\) is \(N=L_1\sin\theta/W\), where \(W\) is the center-to-center distance in the array of acceptor CNTs. Therefore, the total transfer rate is

\begin{equation}
    \Gamma_{12} = \frac{\sin\theta}{\hbar W} \left(\frac{{A_u}^2}{4\pi^2 r_1 r_2} \right)^2 \sum_{s_1,s_2}\sum_{\bm{K}_1}\sum_{\mathscr{M}_2} \frac{e^{-\beta \Omega_{s_1}}}{\mathcal{Z}} \left|J_{\theta}(\bm{K}_1,\bm{K}'_2) \times \tilde{Q}(\bm{K}_1,\bm{K}_2)\right|^2 \left( \frac{dK_2}{d\Omega_{s_2}}\right)_{\Omega_{s_1}}.
\end{equation}

This equation turns out to be divergent at \(\theta \rightarrow 0\). This is simply due to the fact that the transfer rate between completely parallel CNTs is length-independent and we expect the transfer rate from the initial CNT to an infinite number of final CNTs placed at a certain distance to become  infinite. \cite{Lyo2006exciton}

\subsection{The special case of free-exciton transfer between parallel tubes} \label{sec:free_exciton_transfer_between_parallel_tubes}

When the CNTs are very long and parallel to each other, the geometric part of the matrix element yields a Kronecker delta function
\begin{subequations}
    \begin{equation}
        J_\theta(\bm{K}_1,\bm{K}_2) = L \times \delta(K_1,K_2) \times C(\mathscr{M}_1,\mathscr{M}_2;K_1),
    \end{equation}
    \begin{equation}
    \begin{split}
        & C(\mathscr{M}_1,\mathscr{M}_2;K_1)= \frac{2r_1r_2}{A_u^2} \int d\phi \, d\phi' \, e^{2i(\mathscr{M}_2\phi'-\mathscr{M}_1\phi)} \times \cdots \\
        & \hspace{3.5 cm} \cdots \mathcal{K}_0(|2K_1|\sqrt{(r_1\sin \phi-r_2\sin \phi')^2+(D+r_1 \cos\phi - r_2 \cos\phi')^2}),
    \end{split}
    \end{equation}
\end{subequations}
\(L_1=L_2=L\rightarrow \infty \) is the length of CNTs. \(\mathcal{K}_0\) is the modified Bessel function of the second kind.

\noindent Therefore, the matrix element of direct Coulomb interaction becomes

\begin{equation}
    \mathcal{M}_d= \frac{{A_u}^2}{4\pi^2 r_1 r_2} \delta(K_1,K_2) \, C(\mathscr{M}_1,\mathscr{M}_2;K_1) \times \tilde{Q}(\bm{K}_1,\bm{K}_2)
\end{equation}

\noindent Using the conservation of the continuous components of wave vectors \(\bm{K}_1\) and \(\bm{K}_2\) in the matrix element \(\mathcal{M}_d\) in Eq. (\ref{eq:fermi_transfer_rate}), we get

\begin{equation}\label{eq:appendix_aux}
    \begin{aligned}
        \Gamma_{12}
                    = & \frac{2\pi}{\hbar} \left(\frac{{A_u}^2}{4\pi^2 r_1 r_2}\right)^2 \sum_{s_1,s_2}\sum_{\bm{K}_1}\sum_{\mathscr{M}_2} \frac{e^{-\beta \Omega_{s_1}}}{\mathcal{Z}} \left|C(\mathscr{M}_1,\mathscr{M}_2;K_1) \times \tilde{Q}(\bm{K}_1,\bm{K}'_2) \right|^2 \delta(\Omega_{s_1}-\Omega'_{s_2}) \\
                    = & \frac{2\pi}{\hbar}\frac{L}{2\pi} \left(\frac{{A_u}^2}{4\pi^2 r_1 r_2}\right)^2 \sum_{s_1,s_2}\sum_{\mathscr{M}_1}\sum_{\mathscr{M}_2} \int dK_1 \frac{e^{-\beta\Omega_{s_1}}}{\mathcal{Z}} \left|C(\mathscr{M}_1,\mathscr{M}_2;K_1) \tilde{Q}(\bm{K}_1,\bm{K}'_2) \right|^2 \delta(\Omega_{s_1}-\Omega'_{s_2}) \\
                    = & \frac{L}{\hbar} \left(\frac{{A_u}^2}{4\pi^2 r_1 r_2}\right)^2 \sum_{s_1,s_2}\sum_{\bm{K}''_1}\sum_{\mathscr{M}_2} \frac{e^{-\beta\Omega_{s_1}}}{\mathcal{Z}} \left|C(\mathscr{M}_1,\mathscr{M}_2;K''_1) \times \tilde{Q}(\bm{K}''_1,\bm{K}'_2) \right|^2 \left(\frac{dK_1}{d\Omega_{s_1}}\right)_{\Omega''_{s_1}}
    \end{aligned}
\end{equation}

\noindent The primed quantities in the last relation show the excitonic states in the acceptor CNT that conserve the continuous component of the center-of-mass wave vector, i.e., \(K_2=K_1\). The double-primed quantities represent the excitonic states on the donor CNT that conserve both the energy and the continuous part of the center-of-mass momentum in the transfer process, i.e., \(\Omega_{s_1}(\mathscr{M}_1,K_1) = \Omega_{s_2}(\mathscr{M}_2,K_1)\). We should note that the final transfer rate is independent of the CNT length as the partition function, \(\mathcal{Z}\), is linearly dependent on the length of the donor CNT
\begin{equation}
	\begin{split}
			\mathcal{Z} = & \sum_{s_1} \sum_{\bm{K}_1} e^{-\beta \Omega_{s_1}(\bm{K}_1)}	\\
			= & \frac{L}{2\pi} \sum_{s_1} \sum_{\mathscr{M}_1} \int dK_1 ~ e^{-\beta \Omega_{s_1}(\bm{K}_1)},
	\end{split}
\end{equation}

\noindent which cancels the parameter \(L\) in the Eq. (\ref{eq:appendix_aux}), so we obtain
\begin{equation}
    \begin{split}
        \Gamma_{12}
                    = & \frac{2\pi}{\hbar} \left(\frac{{A_u}^2}{4\pi^2 r_1 r_2}\right)^2 \Bigg[ \sum_{s_1} \sum_{\mathscr{M}_1} \int dK_1 ~ e^{-\beta \Omega_{s_1}(\bm{K}_1)}\Bigg]^{-1} \times \\
                    &  \hspace{0.7cm} \Bigg[ \sum_{s_1,s_2}\sum_{\bm{K}''_1}\sum_{\mathscr{M}_2} e^{-\beta\Omega_{s_1}} \left|C(\mathscr{M}_1,\mathscr{M}_2;K''_1) \times \tilde{Q}(\bm{K}''_1,\bm{K}'_2) \right|^2 \left(\frac{dK_1}{d\Omega_{s_1}}\right)_{\Omega''_{s_1}} \Bigg].
    \end{split}
\end{equation}

\end{appendix}

\begin{acknowledgement}

This work was primarily funded by the U.S. Department of Energy (DOE), Office of Basic Energy Sciences, Division of Materials Sciences and Engineering, Physical Behavior of Materials Program, under Award No. DE-SC0008712.
Preliminary efforts (prior to the start of DOE funding) were supported as part of the University of Wisconsin MRSEC, IRG2 (NSF Grant No. 1121288).

\end{acknowledgement}


\begin{mcitethebibliography}{53}
\bibitem[Not()]{Note-2}
Note that in the electron-hole picture, the wave vector of excited electron in
  the valence band is negative of the wave vector of created hole: \(\bm{k}_h =
  -\bm{k}_v\).\relax
\mciteBstWouldAddEndPunctfalse
\mciteSetBstMidEndSepPunct{\mcitedefaultmidpunct}
{}{\mcitedefaultseppunct}\relax
\EndOfBibitem
\bibitem[Not()]{Note-3}
The continuum level is the minimum energy level beyond which the excitonic
  state can essentially be considered as free electron and free hole
  states.\relax
\mciteBstWouldAddEndPunctfalse
\mciteSetBstMidEndSepPunct{\mcitedefaultmidpunct}
{}{\mcitedefaultseppunct}\relax
\EndOfBibitem
\end{mcitethebibliography}
\providecommand{\latin}[1]{#1}
\providecommand*\mcitethebibliography{\thebibliography}
\csname @ifundefined\endcsname{endmcitethebibliography}
  {\let\endmcitethebibliography\endthebibliography}{}

\end{document}